\documentstyle[12pt]{article}

\newcommand{\mysection}[2]{\section[#1]{#2}\setcounter{equation}
{0}}

\newcommand{\be}{\small\begin{equation}}
\newcommand{\ee}{\end{equation}\normalsize\vspace*{-0.1ex}}
\newcommand{\bea}{\small\begin{eqnarray}}

\newcommand{\eea}{\end{eqnarray}\normalsize\vspace*{-0.1ex}}
\newcommand{\bdm}{\small\begin{displaymath}}
\newcommand{\edm}{\end{displaymath}\normalsize\vspace*{-0.1ex}}
\newcommand{\beas}{\small\begin{eqnarray*}}

\newcommand{\eeas}{\end{eqnarray*}\normalsize\vspace*{-0.1ex}}

\newcommand{\n}{\noindent}
\newcommand{\eps}{\epsilon}
\newcommand{\intl}{\int\limits}
\newcommand{\qmu}{\frac{Q^2}{\mu^2}}
\newcommand{\vkmu}{\frac{2 vk}{\mu}}
\newcommand{\mmu}{\frac{m^2}{\mu^2}}

\newcommand{\vslash}{\not\! v}
\newcommand{\proj}{\frac{1+\!\vslash}{2}}
\newcommand{\dd}{{\mbox d}}
\newcommand{\pref}{\frac{C_F}{4\pi N_f}}
\newcommand{\omu}{\left(-\frac{2\omega}{\mu}\right)^{-2 u}}

\begin{document}


\setcounter{page}{0}
\thispagestyle{empty}
\renewcommand{\thefootnote}{\fnsymbol{footnote}}

\vspace*{1.3cm}

\begin{center}
\vspace*{0.5cm}
{\Large\bf
Heavy Quark Effective Theory beyond
Perturbation Theory: Renormalons, the Pole Mass and the
Residual Mass Term
}\\
\vspace{1.4cm}
{\sc M.Beneke} \\
\vspace*{0.3cm} {\it Randall Laboratory of Physics\\
University of Michigan\\ Ann Arbor, Michigan 48109, U.S.A.}\\[0.6cm]
and\\[0.6cm]
{\sc V.M. Braun\footnote{On leave of absence from
St.Petersburg Nuclear Physics Institute, 188350 Gatchina,
Russia}} \\ \vspace*{0.3cm} {\it Max-Planck-Institut
f\"ur Physik\\ -- Werner-Heisenberg-Institut -- \\ D--80805
Munich (Fed. Rep. Germany)}\\[1.4cm]
{\bf Abstract}\\[0.3cm]
\end{center}

\parbox[t]{\textwidth}{
We study the asymptotic behaviour of the perturbative series in
the heavy quark effective theory (HQET)
using the $1/N_f$ expansion.
We find that this theory suffers from an {\it ultraviolet} renormalon
problem, corresponding to a non-Borel-summable behaviour of perturbation
series in large orders, and leading to a principal
nonperturbative ambiguity in its definition. This ambiguity
is related to an {\it infrared} renormalon in the pole mass and can
be understood as the necessity to include the residual mass
term $\delta m$ in the definition of HQET, which must be considered
as ambiguous (and possibly complex), and is required to cancel the
ultraviolet renormalon singularity generated by the perturbative
expansion. The formal status of $\delta m$ is thus identical to
that of condensates in the conventional short-distance expansion
of correlation functions in QCD. The status of
the pole mass of a heavy quark, the operator product expansion for inclusive
decays,  and QCD sum rules
in the HQET are discussed in this context.
}
\vspace*{0.3cm}

\newpage
\renewcommand{\thefootnote}{\arabic{footnote}}
\setcounter{footnote}{0}


\mysection{Introduction}{Introduction}

The past few years have witnessed considerable progress in
understanding the decays of hadrons containing a single heavy
quark in the kinematical regime, where the heavy quark is almost
on-shell. This progress has mainly been achieved through the
implementation of an
effective field theory, which eliminates the heavy quark as a
dynamical degree of freedom \cite{EIC90,GRI90,GEO90}. In the infinite
mass limit, the effective theory unravels new symmetries of
QCD \cite{ISG89}, while providing a systematic approach to treat
corrections to this limit, which are suppressed by inverse powers
of the heavy
quark mass $m_Q$. The number of independent form factors
governing the decays of heavy hadrons is greatly reduced by these
symmetries, which allows to relate the width and spectra of
various decays.
A peculiar property of heavy quark effective theory (HQET)
is that each effective quark field is labelled by the velocity
four-vector $v_\mu$ of the heavy quark, which is conserved
by the strong interactions in the limit of an infinitely heavy
quark. Deep connections have been pointed out \cite{KOR92}
between HQET and the dynamics of Wilson loops [6--11],
related to the infrared behaviour of perturbative QCD and the
factorization of soft divergences.

Given the importance of HQET for different branches of
phenomenology, it is instructive
to investigate its status as a quantum field theory. Thus the
leading order effective Lagrangian,

\be \label{effectivelagrangian}
{\cal L}_{eff}\,=\,\bar{h}_v i v\cdot D h_v + {\cal L}_{light}
\,\ee

\n where $v$ labels the velocity of the heavy quark and
${\cal L}_{light}$ denotes the
Lagrangian for the light degrees of freedom, has been proven
to be renormalizable to all orders in perturbation theory
\cite{BAG94} (see also [7--11]).
 The main objective of this paper is to investigate
the nonperturbative properties of the theory defined by the
Lagrangian in eq.(\ref{effectivelagrangian}), which show up in the
asymptotic behaviour of perturbation theory.
To this purpose we study the structure of singularities of
correlation functions in HQET in the complex plane of the Borel
transform with respect to the strong coupling, an approach that has been
pioneered in \cite{THO77} in its application to QCD.
Our main finding is that apart from the usual sequences
of infrared (IR) and ultraviolet (UV) renormalons, which one expects
to be inherited from QCD, the HQET suffers from an additional series
of UV renormalons, which
are non-Borel-summable. Thus the Lagrangian
${\cal L}_{eff}$ as it stands does not define a respectable
theory, since it is plagued by an {\it ultraviolet} renormalon
problem in the same way as, e.g., QED \cite{LAU77}, albeit for
different reasons\footnote{Ultraviolet renormalons appear in QCD
as well, but since they are Borel-summable in this case, they are
usually not considered as a ``problem".}. In other words,
the high momentum
region of internal integrations in Feynman diagrams renders the
perturbative expansion of Green functions so badly divergent in large
orders as to obstruct their unambiguous definition beyond
perturbation theory.

At this point it is helpful to keep in mind
that despite the sophisticated terminology prevailing the field of
large-order perturbation theory, the physics reflected in the
emergence of renormalons is usually simple and can be understood
without recourse to the asymptotics of perturbation theory. In
QED, for instance, the UV renormalons appear as a disguise of the
good old Landau ghost. Whereas thus the UV renormalons are
presumably fatal for QED as a viable theory (unless it becomes
embedded in a larger nonabelian group), this is of course not so
for the  heavy quark limit of QCD,
since the leading effective Lagrangian, eq.(\ref{effectivelagrangian}),
has to be supplemented by a tower of nonrenormalizable interactions,
suppressed by inverse powers of the heavy quark mass, as well as
renormalization of their coefficients taking into account the
QCD interactions on scales above $m_Q$. It is precisely this
separation of effects on different distance scales into coefficient
functions and matrix elements that introduces {\it infrared}
renormalons into the coefficient functions and {\it ultraviolet}
renormalons into the matrix elements of composite operators at the
{\it same} position in the Borel plane, since the virtual particles
inside the loops do not respect the constraint $k>m_Q$, $k<m_Q$,
respectively, on their internal momentum. This phenomenon is well-known
from the short-distance expansion of correlation functions in
QCD or the $O(N)$-nonlinear sigma model \cite{DAV84}, though to
our knowledge it has never been made explicit in any calculation.
However, this observation alone is not sufficient to cure the UV
renormalon disease in the HQET. The crucial point is that the leading
order effective Lagrangian, eq.(\ref{effectivelagrangian}), extracts
the correct dependence on the heavy quark mass of the Green functions
in full QCD only after subtraction of a term that scales with $m_Q$
(consider, to be definite, the inverse heavy quark propagator as
in \cite{GRI90}). This implies a choice of $m_Q$ that coincides with
the {\it pole} mass of the heavy quark to all orders in perturbation
theory, but arbitrary otherwise. Here the subtleties arise.

In a confining theory like QCD the S-matrix elements
have no poles
corresponding to a physical quark and therefore there is no natural
choice of the expansion parameter $m_Q$. Indeed, the mass of the
lightest meson containing the heavy quark flavour would serve this
purpose as well as any other parameter that differs from the meson
mass by an amount exponentially small in the coupling. This
obvious ambiguity has been known from the very beginnings of HQET and
prompted the authors of \cite{FAL92} to introduce the concept of
a residual mass term, $-\delta m \bar{h}_v h_v$, to be added to the
Lagrangian of eq.(\ref{effectivelagrangian}). The residual mass term,
being of order $\Lambda$, where $\Lambda$ is a characteristic low
energy scale of QCD, will enter the expressions of HQET, once one
leaves the framework of perturbation theory. The predictions of
HQET must be independent of the choice of $m_Q$. Indeed, it has been
shown \cite{FAL92} that the residual mass term enters the matrix
elements of HQET through the combination $\bar{\Lambda}-\delta m$
only, where $m_P-m_Q=\bar{\Lambda}+O(1/m_Q)$ is the difference
between the mass $m_P$ of the meson under consideration and the
heavy quark expansion parameter in the infinite mass limit. This
combination is clearly invariant under the choice of $m_Q$, thus
apparently justifying the choice $\delta m=0$ implicit in most works
on HQET. This conclusion is incorrect. As will be seen in the subsequent
sections, the pole mass develops an IR renormalon, which, when
subtracted in the construction of HQET, {\it necessitates} the
inclusion of a residual mass term as a ``remnant" of this IR
renormalon. If we insist on a formal expansion in $\alpha(m_Q)$ and
$\Lambda/m_Q$, the residual mass term must be considered as
ambiguous (and possibly complex) and this ambiguity is {\it
required} to cancel the UV renormalon in the matrix elements of
operators in the HQET. To express this statement in a different
way: though $\bar{\Lambda}-\delta m$ is invariant under the choice
of $m_Q$, it is {\it not} invariant under the choice of a summation
prescription for the divergent perturbative expansions in the HQET.
We wish to note here that the inherent ambiguity of the quantity
$\bar{\Lambda}-\delta m$
has been conceptually realized in \cite{BIG94},
where some of our results are anticipated. The {\it formal} status of
$\bar{\Lambda}-\delta m$ is thus identical to that of condensates
in the conventional short-distance expansion of correlation functions
in QCD.

This analogy may be pursued further. It has been known for a long time
that the computation of dimensionful parameters like condensates
is a very difficult task for lattice gauge theories, due to
mixing with lower dimensional operators, which manifests itself
in power divergences in the lattice spacing, as the latter is taken to
zero \cite{DIG81,BAN81}. The power divergences require a nonperturbative
regularization procedure, which is essentially equivalent to
fixing a specific summation prescription for the perturbative
expansion. This connects power divergences in lattice gauge theory
to renormalons in the continuum theory, where we might note in
addition that mixing between operators of different dimension
occurs in the continuum precisely through the appearance of
renormalons. Our observation that $\bar{\Lambda}-\delta m$ (and
similar parameters in higher orders of the $1/m_Q$-expansion) is
ambiguous is therefore completely consistent with the observation
of power divergences in the lattice version of HQET that have been
reported in \cite{MAI92}, and which turn out to be an obstacle to
the determination of HQET matrix elements on the lattice starting
at order $1/m_Q$.

The study of the asymptotics of the perturbative series in HQET is a
rather formal subject, but allows to draw several immediate
conclusions of practical importance. The first of them concerns
inclusive $B$-decays, which are receiving much attention presently.
It has been shown \cite{CHA90,BIG92} that nonperturbative
corrections to total inclusive widths can be studied using the
operator product expansion technique, and are suppressed by two powers
of the large $b$-quark mass. It is widely believed that $O(1/m_b)$
corrections to the total widths are absent, provided the latter are
expressed in terms of the pole $b$-quark mass, see \cite{MAN94} for the
clearest presentation of
this conviction. The nonperturbative $1/m^2_b$ corrections turn out to
be very small in reality, and this result has already triggered attempts
to determine the $b$-quark pole mass from the data on the total decay
rates \cite{LUK94}. Our results show that the absence of $1/m_b$
corrections is illusory. Different prescriptions for the summation of
the perturbative series defining the pole mass introduce a principle
uncertainty in the predictions for the decay rates. The data on the
total inclusive widths can not be used to {\em determine} the pole
mass, but rather to {\em define} it. This definition is not worse, but
also not better than any other {\it phenomenological} definition, e.g.
from the QCD sum rules for the $B$-mesons or $\Upsilon$ family, see
 \cite{VOL80,REI85,NAR86}. The existing estimates for the value of the
$b$-quark pole mass fall in the range $m_b = 4.55-4.85$ GeV, indicating
that possible uncertainty is of order a few hundred MeV.
 We find very similar values
for the intrinsic uncertainty in the pole
mass  from our results on the ambiguity in the summation of the
perturbative series.

Second, we address the QCD sum rule technique \cite{SHI79},
 which has been used to
obtain quantitative estimates for $\bar\Lambda$ and other observables
in HQET (see \cite{NEU93} for a review).
 Our analysis suggests that the residual mass term should
be included as an additional phenomenological parameter (like the
condensates) into the expansion of correlation functions in the HQET.
The effect of this parameter is, however, very specific.
We show that, loosely speaking, the renormalons associated with
the residual mass term can be ``summed up" and result in an
ambiguity in the momentum scale of the correlation function, so that
quantities like the $B$-meson
decay constant $f_B$ and the heavy quark kinetic energy
\cite{BAL93} are formally not affected.

The most important question is whether this ambiguity is important
{\em numerically}.
Again, we might
appeal to the more familiar situation of the short-distance
expansion of correlation
functions in QCD, where the gluon condensate, e.g., has been
determined despite its ambiguity, observing that its actual value
is ``large" in the sense that the IR renormalon of perturbation
theory may be ignored. It is however very important to recall
that this can be justified only {\it a posteriori} by the very
success of the sum rules. There is no guarantee that the same
conclusion applies to the parameters of the heavy quark expansion.

The further presentation is organized as follows.
In Sect. 2 we find it useful to recapitulate some facts
on the operator product expansion and IR renormalons of the
correlation function of light quark currents. This will also
allow us to introduce the basic notions in dealing with large-order
perturbation theory that will be needed later on. In Sect. 3 we
study in detail the perturbative expansion of the inverse
propagator of a massive quark in QCD and its matching onto the
heavy quark expansion. New IR and UV renormalons will be seen
to emerge in this limit. Calculations to all orders in
perturbation theory are performed in an expansion in $1/N_f$,
with $N_f$ the number of light flavours. We move to the
consideration of the correlation function of heavy-light currents
in Sect. 4, where the analytic properties of its Borel-transform
are obtained and discussed.
Sect. 5 is devoted to applications of our results to the practice
of QCD sum rule calculations and in Sect. 6 we present a summary and
conclusions.

Two appendices deal with some more technical issues. In App. A
we show, how the renormalization of the Borel transform proceeds
in the leading order of the $1/N_f$ expansion. For simplicity
of notations, the explicit derivation is given for the self-energy
of the heavy quark. In App. B we compute the scalar two-loop
integral with arbitrary power of the propagator of the interchanged
scalar. The singularity structure of this integral is required
to obtain the analytic structure of the Borel-transformed
correlation function discussed in Sect. 4.



\mysection{IR renormalons and the short-distance expansion of light
quark current correlation functions}{IR renormalons and the
short-distance expansion of light quark current correlation functions}

The best studied (see e.g. \cite{MUE92,ZAK92} for two recent
expositions) and most transparent quantity to exhibit the relation
of the IR asymptotics of perturbation series to the short-distance
expansion is provided by the correlation function

\be
\Pi_{\mu\nu}(q) = (q_\mu q_\nu-q^2 g_{\mu\nu})\,\Pi(Q^2) =
i\int\mbox{d}^4 x\,e^{i q x}\,\langle 0|T\{j_\mu^\dagger(x) j_\nu(0)\}| 0
\rangle \qquad Q^2=-q^2
\ee

\n of two vector currents $j_\mu(x)=\bar{q}(x)\gamma_\mu q(x)$ of
light, i.e. massless, quarks. It is useful to recall this relation
in detail, since the same concepts will recur in the more
intricate context of the heavy quark expansion. We hope that the
yet inevitable sketchiness of our presentation does not appall
the more rigorous minded readers.

Let us first focus on the perturbative expansion of $\Pi(Q^2/\mu^2,
\alpha(\mu))$ in the strong coupling. One may easily identify
one source of divergence of this expansion in large orders. To this
end, one investigates the diagram of Fig. 1 with the gluon line
dressed by a chain of fermion, gluon and ghost loops, summation of
which is essentially equivalent to placing the running coupling
$g(k)$ at the vertices, where $k$ is the momentum which flows through
the gluon line\footnote{This statement is strictly true only in
QED. In QCD, dressing of a gluon by a chain of bubbles is not a
gauge-invariant procedure and one must leave the framework of bubble
summation to obtain the correct coefficient $\beta_0$ in front of the
logarithm in eq.(\ref{irintegral}).}. Since we are interested in
the contribution from soft gluons, $k\ll Q$, after performing all
other integrations, we expand in $k^2/Q^2$ and obtain

\be\label{irintegral}
\Pi\left(\qmu,\alpha(\mu)\right)\,\ni\,\intl^\lambda\frac{
\mbox{d}k^2}{k^2} \left(\frac{k^2}{Q^2}\right)^m
\alpha(k) = \sum_n \alpha(Q)^{n+1}\intl^\lambda\frac{
\mbox{d}k^2}{k^2} \left(\frac{k^2}{Q^2}\right)^m
\left(\beta_0\ln\frac{k^2}{Q^2}\right)^n\,,
\ee

\n where $m$ is a natural number, ``$\ni$" denotes a
contribution to the asymptotic behaviour, which need not be the
dominant one, and $\lambda$ regularizes the UV divergence
introduced through the expansion in $k^2/Q^2$. The first
coefficient of the $\beta$-function, $\beta_0$, is negative in
our definition.
The logarithmic behaviour of the running coupling
drives the gluon line to increasingly softer momentum, $k\sim
Q e^{-n/(2 m)}$, as $n$ becomes large. At the same time, the logarithm
is large in this regime, no matter what (fixed!) renormalization
scale one chooses (we have taken $\mu=Q$ for convenience). As a
result a factorial divergence of the perturbative expansion

\be\label{irasymptotics}
\Pi\left(\qmu,\alpha(\mu)\right)\,\equiv\,\sum_n \Pi_n\left(
\qmu\right) \alpha(\mu)^{n+1}\,\ni\,\sum_n \left(-\frac{\beta_0}
{m}\right)^n n! \,\alpha(Q)^{n+1}
\ee

\n arises\footnote{For completeness, we note that a similar, but
sign-alternating divergence proportional to $(\beta_0/m)^n n!$ is
present by the same reasoning, applied to the ultraviolet region,
$k\gg Q$.}. We may still make progress and define the Borel-transform
of $\Pi$ by

\be\label{defboreltransform}
\tilde{\Pi}\left(\qmu,t\right)\equiv\sum_n \Pi_n\left(\qmu\right)\,
\frac{t^n}{n!}\,.
\ee

\n In favourable circumstances \cite{SOK80}, $\Pi$ can be recovered
despite its divergent expansion from the integral representation

\be\label{inverseborel}
\Pi\left(\qmu,\alpha(\mu)\right) = \intl_0^\infty \mbox{d} t\,
e^{-t/\alpha(\mu)}\,\tilde{\Pi}\left(\qmu,t\right)\,.
\ee

\n However, from eq.(\ref{irasymptotics}) one infers that the Borel
transform $\tilde{\Pi}$ has (IR renormalon) singularities at
$t=-m/\beta_0$ on the integration contour. The naive Borel summation
fails in QCD and does not define $\Pi$ unambiguously. As a measure
of this ambiguity one may take the difference between the contour
prescriptions above and under the real axis in the complex Borel plane.
One then concludes that within perturbation theory we can account
for the infrared domain only up to terms of order

\be\label{uncertainty}
\delta\Pi_{IR}(Q^2)\,\ni\,\exp\left(\frac{m}{\beta_0\alpha(Q)}\right)
\sim\left(\frac{\Lambda^2}{Q^2}\right)^m\,.
\ee

\n This deficiency of pertubation theory has a profound reason. In the
real world, quarks and gluons are confined into hadrons and one does
not expect this phenomenon to occur as a mere consequence
of summation of the perturbative series and analytic continuation
to the physical region. If QCD is to have any significance for
the real world, perturbation theory must be incomplete and the IR
renormalons are just a reminder that nonperturbative terms must be
added. Moreover, the location of singularities on the positive real
Borel axis traces the order of magnitude of these effects.

In case of correlation functions at deep euclidean momentum, the
framework for a systematic incorporation of nonperturbative effects
is the short-distance expansion (SDE) of the operator product
$j_\mu(x) j_\nu(0)$, which leads to the improved expression
\cite{SHI79}

\be\label{sde}
\Pi\left(\qmu,\alpha(\mu)\right)\,=\,\sum_n p_n\left(
\qmu\right) \alpha(\mu)^{n+1} + C_{GG}\left(\qmu,\alpha(\mu)\right)
\frac{1}{Q^4} \langle 0|\frac{\alpha}{\pi} G G|0\rangle (\mu)
+ O\left(\frac{1}{Q^6}\right)\,,
\ee

\n where the new input is given by the gluon condensate and its
Wilson coefficient function and higher power-suppressed terms
involve the vacuum expectation value of higher dimensional
operators. This representation is supposed to yield an
{\it unambiguous} answer for $\Pi$, including all nonperturbative
effects up to terms of order $1/Q^6$. How is this possible in view
of the above ambiguities inherent to the perturbative contribution
due to IR renormalons? This question can be answered from a formal
and a practical position  and we choose to begin with the first.

The nonperturbative definition of condensates is indeed a delicate
problem. We may pick a physical cutoff, in which case the operator
$\alpha/\pi GG$ can mix with lower dimensional ones, the unit
operator in particular. In order to define the normal product, one
must subtract these contributions, which is an ambiguous procedure
due to the occurence of renormalons in their series in the limit,
when the cutoff is removed. To fix an
exponentially small in $\alpha$ term like $\langle 0|\alpha/\pi GG
|0\rangle$, one must first give a meaning to the divergent perturbative
expansions in lower dimensional operators, which may be complex,
depending on the contour prescription for the singularities in their
Borel representation. Without this specification, we are thus led to
the notion of ambiguous (and possibly complex) condensates\footnote{
A beautiful illustration of this phenomenon has been given in
\cite{DAV84} within the $1/N$-expansion of the $O(N)$-nonlinear
sigma model. Within this expansion dimensional regularization
provides a nonperturbative regularization method. Power divergences
are then seen to appear as poles at dimensions depending on the
order of perturbation theory. To all orders, they accumulate at
$d=4$, forcing the limit $d\rightarrow 4$ to be taken through the
upper or lower complex $d$-plane, with a different (and complex)
result, depending on how the limit is taken.}, although by no means
this implies that the correlation function $\Pi$ is ambiguous (or
complex). To the contrary, this imaginary part is correlated with
the summation prescription for the IR renormalon divergence in such
a way, that the final answer for $\Pi$ is real and unambiguous.
There are two messages to be taken from these considerations:
First, perturbation theory ``knows" about nonperturbative effects
through the singularities of the Borel transform on the positive
axis\footnote{Clearly, perturbation theory does not know about
{\it all} nonperturbative effects. For example, in the finite mass
case a quark condensate term, $m\langle\bar{q} q\rangle$, appears
in eq.(\ref{sde}), which is not seen as a renormalon, because
the operator $\bar{q} q$ can not mix with lower dimensional ones,
owing to its transformation properties under chiral symmetry. We
shall check this explicitly in Sect. 4.}. Second, once these effects
are identified, they yield strong constraints for the nature of
the IR renormalon singularities \cite{PAR79,MUE85,BEN93a}. Thus,
from the absence of a dimension-two condensate in eq.(\ref{sde})
one excludes the existence of a singularity at $t=-1/\beta_0$,
which according to eq.(\ref{uncertainty}) would call for a
$1/Q^2$-term. In addition, the condensates (and therefore their
ambiguity) satisfy renormalization group equations, which determine
the $\alpha$-dependence of the ambiguity. This must match the
ambiguity in the Borel representation of the perturbative series,
thus fixing the nature of the corresponding singularity.

We want to emphasize that the appearance of imaginary parts in
exponentially small components added to a divergent series
is far more general than the SDE, and is just an example of the
so-called Stokes-phenomenon \cite{DIN73},
which generically arises in asymptotic
expansions with fixed-sign divergence. It is important to realize
that the Stokes discontinuities are {\it formal}: After proper
summation of all terms, one obtains an analytic function, and,
paradoxically, the Stokes discontinuities occur, {\it because}
the function, which is represented by the asymptotic expansion,
is analytic. It is the most economic way for an asymptotic expansion
to keep up with the analyticity of the function over a finite
phase range of the expansion parameter.

Miraculously, the formal complexities, which we have just reviewed,
have never been an obstacle to the practice of QCD sum rules, where,
for instance, the gluon condensate is added with some definite value
to a few low-order terms of the perturbative expansion. To
understand this better, we observe that, although eq.(\ref{sde}) gives
the correct asymptotic expansion of $\Pi$, it is not quite the
implementation of Wilson's operator product expansion program. This
is {\it not} designed to separate perturbative and nonperturbative
effects into coefficient functions and matrix elements, respectively,
an intrinsically ambiguous procedure. Instead it disentangles
the physics on different distance scales. Thus one should introduce
the factorization scale $\mu<Q$ properly, i.e. cut out the region
$k<\mu$ from the momentum integrations in the Feynman diagrams
contributing to the coefficient functions and attribute it to the
condensates as a non-universal piece. Although this is extremely
awkward in practice -- see \cite{NOV84,BEN93b} for illustrative
examples --, one may guess conceptually, how eq.(\ref{sde}) is
modified. The first perturbative coefficients are not significantly
affected, because they are contributed by internal momenta
$k\sim Q$. As one progresses towards higher orders, there is a
factorially large contribution from momenta $k\sim Q e^{-n}$,
which eventually is eliminated by the constraint $k>\mu$ on the
internal integrations. The IR renormalons disappear from all
Wilson coefficients. In turn the condensates develop a complicated
dependence on $\alpha$. An asymptotic expansion in $\alpha$ reveals
the IR renormalon as a perturbative contribution to, e.g., the
gluon condensate\footnote{The asymptotic behaviour of the
perturbative contribution {\it is} universal. To connect to the
formal position, note that $c^\prime$ should formally be considered
ambiguous and carries the Stokes discontinuity.}:

\be
\langle 0|\frac{\alpha}{\pi} G G|0\rangle (\mu) = c \mu^4
\sum_n \left(-\frac{\beta_0}{2}\right)^n n!\, n^{-2\beta_1/
\beta_0^2}\,\alpha(\mu)^{n+1} + c^\prime \mu^4 e^{2/(\beta_0
\alpha(\mu))} \alpha^{2\beta_1/\beta_0^2}\,(1+O(\alpha))
\ee

\n The whole point of the QCD sum rules relies on the fact that this
perturbative contribution is small compared  to ``anomalously" large,
genuine nonperturbative effects in the infrared, and can be neglected
\cite{SHI93}. From the theory point of view this ``rule of discarding the
perturbative piece of condensates" remains one of the mysteries
of QCD. It could not have been guessed in advance and is justified
only by the empirical fact that the sum rules work. In particular,
it is far from obvious that the IR renormalons are irrelevant
outside the context of the SDE.

Though the existence of renormalons can hardly be doubted on
physical grounds, a literal proof does not exist even for
the scalar $\Phi^4$
theory in four dimensions\footnote{The intrepid reader is referred
to ref. \cite{DAV88}, which comes closest to a proof.} due to the
failure of continuum field theory in providing a nonperturbative
definition of the theory. For this reason, various forms of
$1/N$ expansions have become the state-of-the-art approach to
renormalons. In QED and, for lack of anything more appropriate, also
in QCD, one chooses $1/N_f$ as an expansion parameter, where
$N_f$ is the number of massless fermions. To organize this
expansion, define $a=\alpha N_f$ and expand in $1/N_f$ at fixed
$a$. In order $1/N_f$, where the renormalons appear first, one has
to calculate all diagrams with an arbitrary number of fermion
loops inserted into the gluon line of the two-loop diagrams such
as in Fig. 1. Since all the dependence on the order in $a$ resides
in the number of fermion bubbles, the summation of these diagrams
can be taken directly on the gluon propagator, see Fig. 2. The
(untruncated) sum of $n$ bubbles is given by

\be
D_{\mu\nu,n}^{AB}(k) = i\delta^{AB} \frac{
k_\mu k_\nu - k^2 g_{\mu\nu}}{(k^2)^2}\,(-\pi_0(k^2))^n
\ee

\n where the Landau gauge has been assumed and renormalization of
the fermion bubbles is already understood. Thus

\be
\pi_0(k^2) = -\frac{a}{6\pi} \left(\ln\frac{-k^2}{\mu^2} + C\right)\,
\ee

\n with a scheme-dependent finite renormalization constant $C$.
In the $\overline{MS}$-scheme $C=-5/3$. It is then
easy to find that the {\it Borel-transformed} correlation function
to order $1/N_f$ is simply obtained by replacing the usual
gluon propagator by

\be\label{gluonprop}
D_{\mu\nu}^{AB}(k,u) = \sum_{n=0}^\infty \frac{1}{n!}\,
D_{\mu\nu,n}^{AB}(k)\,\left(\frac{t}{a}\right)^n =
i\delta^{AB} \left(\frac{e^C}{\mu^2}\right)^{-u} \frac{
k_\mu k_\nu - k^2 g_{\mu\nu}}{(-k^2)^{2+u}}\,.
\ee

\n We have defined $u\equiv -\beta_0 t$ with $t$ the Borel
parameter. This propagator includes the
renormalization of the fermion bubbles, which is equivalent
to renormalization of the coupling in the exponent of
eq.(\ref{inverseborel}). In this order of the
flavour expansion, gluons do not contribute to the $\beta$-function
and $\beta_0=1/(6\pi)$. Unfortunately, we lost the asymptotic freedom
property and QCD is identical to QED to this order! In particular,
the IR renormalons move to the negative real axis in the Borel plane.
Despite its obvious inadequacy to describe the dynamics of QCD, the
$1/N_f$ expansion has nonetheless proven successful in detecting
the position of renormalons, once we substitute for $\beta_0$ its
full value $\beta_0=1/(6\pi)-11/(4\pi N_f)$. The reason is of
course the intimate relation of renormalons to the scale
dependence of the effective coupling. Thus, tracing the fermionic
contribution to the $\beta$-function, we get the remaining part
-- i.e. the gluon and ghost bubbles and whatever else is needed
to restore gauge invariance -- for free. As an illustration
consider  the remarkably simple expression for the Borel transform
of the correlation function of two vector currents to order
$1/N_f$ \cite{BEN93c}\footnote{Compared to ref. \cite{BEN93c},
the sign in the definition of $u$ has been changed and the overall
coefficient adjusted to the QCD case.} (see also \cite{BRO93}):

\be
\tilde{\Pi}\left(\qmu,u\right)\,=\,-\frac{8}{3\pi^3 N_f}
\left(\qmu e^C\right)^{-u} \frac{1}{1-(1-u)^2} \sum_{k=2}^
\infty \frac{(-1)^k k}{(k^2-(1-u)^2)^2}
\ee

\n It exhibits the expected UV renormalons at negative integers
(the singularity at $u=0$ must be killed by renormalization or
by taking one derivative with
respect to $Q^2$) and the IR renormalons
at $u=2,3,\ldots$. As required there is no IR renormalon at $u=1$,
i.e. $t=-1/\beta_0$, which would correspond to a dimension-2 operator
in the SDE, and the IR renormalon at $t=-2/\beta_0$ can be shown
to be a simple pole as a consequence of the vanishing one-loop
anomalous dimension of the gluon operator $\alpha/\pi GG$
\cite{ZAK92,BEN93c}. The $1/N_f$ expansion can not detect all
singularities of the Borel transform that should be present in
QCD. Instanton singularities produce effects that scale as
$\exp(-4\pi N_f/a)$ and will not be seen to any order in $1/N_f$.
As far as renormalons are concerned, however, all present
knowledge supports the assumption that the $1/N_f$ expansion is
relevant, provided we substitute $\beta_0$ by its full value.

We will employ the $1/N_f$ expansion in the ensueing sections
because of its transparency in displaying directly the
singularities in the Borel plane, but wish to stress again
that this expansion does not contain more information than what
can already be extracted from an asymptotic expansion of the
Feynman integrands of the low-order diagrams. Indeed, this is
just the way to obtain the coefficient functions of higher
dimensional operators in the SDE.



\mysection{The heavy quark expansion: Matching to all orders}
{The heavy quark expansion: Matching to all orders}

The starting point for HQET, which may be borrowed from
nonrelativistic QED, is that the heavy quark spinor splits
into a large and a small component, when the heavy quark is almost
on-shell. One therefore introduces an effective heavy quark
field

\be h_v(x) = \proj\,e^{i m_Q (v\cdot x)}\,Q(x)
\ee

\n by projecting on the large component and removing a
phase. $v$ is the four-velocity of the heavy quark and $m_Q$ is
usually referred to as the ``heavy quark mass". In this way one
arrives at the effective Lagrangian eq.(\ref{effectivelagrangian}).
The effective propagator reads

\be \frac{1+\vslash}{2}\,\frac{i}{v\cdot k}\ee

\n and the quark-gluon vertex is $-ig v_\mu T^A_{ab}$ (provided one
multiplies by $(1+\!\vslash)/2$ for each external heavy quark line),
which reveals immediately the flavour and spin independence of
the effective theory.

In the following subsections we study in detail the heavy
quark expansion of the inverse propagator in QCD to all
orders in perturbation theory, its matching
onto the HQET, the pole mass
of the heavy quark in QCD and the role of the expansion parameter $m_Q$.
As it turns out, the inverse propagator is not only the simplest,
but also the most instructive object to consider. We consider
the theory with one heavy and $N_f$ massless quarks and expand
in $1/N_f$.

\subsection{The self-energy of a heavy quark}

The full propagator in the effective theory can be written as

\be
\proj\,i S_{eff}(vk)\qquad S_{eff}^{-1}(v k)\equiv vk -
\Sigma_{eff}(vk)\,.
\ee

\n The Borel transform of the self-energy is obtained from the
diagram depicted in Fig. 3, where the gluon line represents
the summation over an arbitrary number of renormalized
fermion bubbles as
explained in Sect. 2. Using eq.(\ref{gluonprop}), we are left
with a single integration over the gluon momentum with the
result ($C_F=4/3$)

\be\label{partrenselfenergy}
\tilde{\Sigma}^{part.ren}_{eff}(v k,u) = \pref\,v k \left(
-\vkmu\right)^{-2 u} e^{-u C}\,(-6)\,\frac{\Gamma(-1+2 u)
\Gamma(1-u)}{\Gamma(2+u)}\,.
\ee

\n All calculations have been performed in dimensional regularization. It
turns out that at generic $u$ the result is finite and one can
actually put $d=4$ as done in eq.(\ref{partrenselfenergy}). The
only renormalization that has still to be done is to account for the
overall subtraction of the whole diagram. As shown in App. A,
this simply amounts to subtracting the pole of the Borel transform
at $u=0$ and eq.(\ref{partrenselfenergy}) is corrected to

\be\label{heavyself}
\tilde{\Sigma}_{eff}(v k,u) = \tilde{\Sigma}^{part.ren}_{eff}
(v k,u) + \pref\,v k\left(-\frac{3}{u} + R_{\Sigma_{eff}}(u)
\right)\,.
\ee

\n The function $R_{\Sigma_{eff}}(u)$ is entire in the Borel plane,
if a renormalization scheme with analytic counterterms is chosen
(such as $\overline{MS}$) and can be neglected in the discussion
of singularities. More on the issue of scheme-dependence
can be found in App. A. From the definition of the Borel transform in
eq.(\ref{defboreltransform}) one can read off that the coefficient
$p_n$ of the perturbative expansion
of the self-energy can be recovered from an expansion of the Borel transform
in $u$. More precisely, to obtain
the coefficient of $a^{n+1}$ in the expansion in the coupling,
one has to take $n$ derivatives at $u=0$:

\be p_n\left(\frac{v k}{\mu}\right) = (-\beta_0)^n\,
\frac{\mbox{d}^n}{\dd u^n}\tilde{\Sigma}_{eff}(v k,u)\big|_{u=0}
\ee

\n In particular, the large-$n$ behaviour is dominated by the pole
closest to the origin $u=0$ of the Borel plane.

Let us now scrutinize the singularities of
$\tilde{\Sigma}_{eff}(v k,u)$. We  find IR renormalons at positive
integer $u$ (i.e. $t=-n/\beta_0$ on
the positive Borel axis\footnote{As mentioned previously,
we always abstract from the $1/N_f$ expansion
and restore the full $\beta_0$, which is negative.}) and UV
renormalons at $u=1/2,-1/2,-1,-3/2,-5/2,\ldots$. To ascertain the
UV or IR nature of a given singularity, one either has to inspect
the diagram explicitly or to observe the general rule that whenever
$u$ occurs with a positive sign in the argument of the Gamma-function
in the numerator of a Borel transform, it is UV and with a negative
sign it is IR. We are hardly surprized to find IR renormalons on
the positive axis, since the effective theory must coincide with
QCD in the infrared. The disturbing novelty is an {\it ultraviolet}
renormalon at {\it positive} $u=1/2$,
which is {\it not} Borel-summable and indicates an
intrinsic nonperturbative ambiguity of HQET that can not be
remedied by any nonperturbative effect. The Lagrangian
eq.(\ref{effectivelagrangian}) {\it as it stands} must therefore be
abandoned as a
sensible quantum field theory beyond perturbation theory.

It is easy to clarify the origin of this UV-renormalon. The first order
correction to the self-energy is proportional to

\be
\int\frac{\dd^4 p}{(2\pi)^4}\,\frac{1}{p^2 (vp+vk)}\,,
\ee

\n which is {\it linearly} divergent. The emergence of a linear
divergence at this point is physically very transparent. A very
heavy quark interacts with its environment only as a static, point-like
colour source. The self-energy is then simply given by the
energy of the Coulomb field of the source, $\alpha(r)/r$, which
is linearly divergent for a point-like object. The divergent
part must be included into the renormalization of the mass of
the source.
As a consequence of this linear divergence, the
series of UV renormalons starts from $u=1/2$, extending
to $u=-\infty$, and not from $u=0$ to $u=-\infty$, as usual. Whereas
the standard (dimensional)
renormalization of logarithmic divergences subtracts
the pole at $u=0$, it does not subtract the linear divergences. This
procedure is legitimate as long as one stays within perturbation theory,
where a distinction between powers and logarithms is meaningful.
Beyond perturbation theory the linear divergences can not be ignored. One
could therefore think of introducing a physical, dimensionful
cutoff $\lambda$. Inevitably, one induces a coun\-ter\-term $\lambda
\bar{h}_v h_v$, which can not be absorbed into the parameters of the
effective Lagrangian, eq.(\ref{effectivelagrangian}). This
reasoning suggests that the HQET may be rescued at the price of
introducing an additional parameter that appears as a mass term.
Note the similarity with massless $\Phi^4$ theory in four
dimensions. We encounter a similar kind of fine-tuning, which is
very familiar from the scalar theory, in HQET, where the natural mass
of the {\it effective} heavy quark is $m_Q$, the UV cutoff of HQET.
This will spoil the heavy
quark expansion, since the Lagrangian in eq.(\ref{effectivelagrangian})
has been constructed precisely to eliminate the $m_Q$-dependence.
To avoid this problem one must impose a renormalization condition
on the two-point function that fixes the mass to zero, which is
automatically achieved by dimensional renormalization. This does
not prevent the appearance of a mass term beyond perturbation
theory and,
indeed, this is what occurs automatically, when the heavy
quark limit is constructed with an expansion parameter which is well-defined
beyond perturbation theory.

\subsection{Renormalon singularities in the pole mass}

To explain our previous assertion, we digress in this subsection
from HQET and deal with the pole mass in QCD. To this end, consider
the self-energy of a massive quark. The full propagator
is defined by

\bea\label{fullprop}
i S(p,m) &=& \frac{i}{\not\!p - m - \Sigma(p,m)}\\
\Sigma(p,m) &=& m\,\Sigma_1(p^2,m) + (\not\!p -m)\,\Sigma_2(p^2,m)
\nonumber\eea

\n The diagram analogous to the one in Fig. 3 but with a quark of
finite mass yields the Borel transforms:

\bea\label{massiveselfenergies}
\tilde{\Sigma}_1(p^2,m,u) &=& \pref\left(\mmu\right)^{-u} e^{-uC}\,
3\,\Gamma(1-u)\Gamma(u)\,{}_2 F_1\left(u,1+u,2;\frac{p^2}
{m^2}\right) \nonumber\\
&& + \,\tilde{\Sigma}_2(p^2,m,u) + \pref\left(
-\frac{3}{u}+R_{\Sigma_1}(u)-R_{\Sigma_2}(u)\right)\\
\tilde{\Sigma}_2(p^2,m,u) &=& \pref\left(\mmu\right)^{-u} e^{-uC}
\left(-\frac{3}{2}\right) u\,
\Gamma(1-u)\Gamma(u)\,{}_2 F_1\left(u,2+u,3;\frac{p^2}
{m^2}\right)\nonumber\\
&& + \,\pref\,R_{\Sigma_2}(u)\nonumber
\eea

\n Here $m$ denotes the renormalized mass (in the scheme specified by
$C$ and the functions $R_{\Sigma_1}(u)$ and $R_{\Sigma_2}(u)$)
at the normalization point $\mu$ and ${}_2 F_1$ is the hypergeometric
function. Let us pause for a glance at the singularities of the
self-energy. If the potential singular point $p^2=m^2$ of the
hypergeometric function is avoided, the UV renormalons occur at
negative integers and the IR renormalons at positive integers, just as
expected in QCD from the considerations of Sect. 2. The IR renormalon
at $u=1$ is not in conflict with the short-distance expansion, which
for the inverse propagator contains gauge-variant operators of
dimension two like $A_\mu A_\mu$, where $A_\mu$ is the gluon field.

Next we move to the pole mass, which not only is the key quantity in the
derivation of the HQET, but also has a considerable interest in
itself, as it appears in many phenomenological applications. Then it is
important to keep in mind that the concept of a pole mass has no
natural extension beyond perturbation theory\footnote{There might still
be a pole in the propagator, when it is defined in a nonperturbative way.
Corresponding to a coloured object, it is however alien to our world.}.
Thus we have to
find the solution to

\be
\not\!p - m - \Sigma(p,m)\big|_{p^2=m^2_{pole}} = 0
\ee

\n in the form of a series expansion

\be\label{polemassseries}
m_{pole}\left(\frac{m}{\mu},a\right) = m\left(1 + \sum_{n=0}^\infty
r_n\left(\frac{m}{\mu}\right) a^{n+1}\right)\,.
\ee

\n Keeping in mind that the self-energy is of order $1/N_f$, we find

\be m_{pole} = m\left(1 + \Sigma_1(m^2_{pole},m) +
O\left(\frac{1}{N_f^2}\right)\right)\,.
\ee

\n This is still a complicated implicit equation for $m_{pole}$. A
crucial  simplification arises from the observation that
$m_{pole} = m + O(1/N_f)$, which eliminates $m_{pole}$ from the r.h.s.
to order $1/N_f$.
Taking the Borel transform of eq.(\ref{polemassseries}) and the
explicit expression eq.(\ref{massiveselfenergies}) for $\Sigma_1$, we
obtain\footnote{If a constant term is present, it is useful to
include it into the Borel-transform with a $\delta$-function, which
preserves the form of the inverse Borel transform,
eq.(\ref{inverseborel}).}

\be\label{polemass}
\tilde{m}_{pole}\!\left(\frac{m}{\mu},u\right) = m \left(\delta(u)
+ \pref\left[
\left(\mmu\right)^{-u} \!\!e^{-u C}\,6\,(1-u)\,\frac{
\Gamma(u)\Gamma(1-2 u)}{\Gamma(3-u)} - \frac{3}{u} +
R_{\Sigma_1}(u)\right]\right)\,.
\ee

\n The scheme dependence residing in $m$ cancels the
scheme dependence  of the expression in
brackets up to terms of order $1/N_f^2$, and $m_{pole}$ proves to be
scheme-invariant, as it must be. In the $\overline{MS}$-scheme
one finds (following the procedure of App. A)
$R_{\Sigma_1}(u)=-5/2+35 u/24+O(u^2)$ and

\be r_0^{\overline{MS}}\left(\frac{m}{\mu}\right) =
m^{-1}\times\tilde{m}_{pole}\left(
\frac{m_{\overline{MS}}}{\mu},
u=0\right)= \frac{C_F}{4\pi N_f}
\left[4 + 3\ln\frac{\mu^2}{m^2_{\overline{MS}}}\right]
\ee

\n reproduces the well-known relation between the pole mass and the
$\overline{MS}$-mass to lowest order ($r_0$ is the coefficient of
$a=\alpha N_f$).

It is seen immediately from eq.(\ref{polemass}) that
the on-shell limit created new
singularities in the Borel transform! The pole mass has an {\it
infrared} renormalon at $u=1/2$, implying a stronger divergence
of the series, eq.(\ref{polemassseries}), than
for the expansion of the self-energy at the
non-singular points $p^2\not=m^2$. Without any reference
to HQET this tells us, that the pole mass can only be defined up
to terms of order $\Lambda_{QCD}$, unless some {\it ad hoc}
definition is employed\footnote{Or, to make contact with one
of our previous footnotes: The pole of the nonperturbatively
defined propagator can be obtained from eq.(\ref{polemassseries})
only by adding terms proportional to $\Lambda_{QCD}$.}.
To make this precise, one may attempt
Borel summation and take half the difference of the values obtained
from the contour prescription above and below the IR renormalon pole as a
measure of the inherent uncertainty of the pole mass,

\be
\delta m_{pole} = \left|\frac{1}{2}\intl_{C^\prime}\dd t\,e^{-t/a(\mu)}\,
\tilde{m}_{pole}
\left(\frac{m}{\mu},-\beta_0 t\right)\right|\,,
\ee

\n where the contour $C^\prime$ wraps around the positive real axis
with the origin excluded. This results in

\bea
\delta m_{pole} \!&=& \!\frac{C_F}{2 N_f |\beta_0|}\,e^{-C/2}\,m(\mu=m)
\exp\left(\frac{1}{2 N_f\beta_0\alpha(m)}\right) \\
\!&=&\! \frac{C_F}{2 N_f |\beta_0|}\,e^{-C/2}\,\Lambda_{QCD}\left(
\ln\frac{m^2}{\Lambda_{QCD}^2}\right)^{\beta_1/(2\beta_0^2)}
\,,\nonumber
\eea

\n where $\beta_1=-1/(4\pi N_f)^2\times (102-38 N_f/3)$ is the second
coefficient of the $\beta$-function (for the rescaled coupling
$a=\alpha N_f$) and we have
indicated the renormalization point explicitly. Note
that $e^{-C/2}\,\Lambda_{QCD}$ is scheme-independent
\cite{CEL79,BEN92} and the remaining scheme-dependence is suppressed
by $1/\ln (m^2/\Lambda_{QCD}^2)$. An alternative (but
scheme-dependent) estimate of $\delta m_{pole}$ can be obtained
from the minimal term of the perturbative expansion and differs from
the above by a factor $(4 N_f|\beta_0| \alpha(m)/\pi)^{1/2}\approx
0.5$. For a numerical estimate we use an average and obtain

\be \label{poleambiguity}
\delta m_{pole}\approx (170 - 280)\,\mbox{MeV}\,.
\ee

\n The numerical values are given for the bottom quark and
four light flavours. We have varied $\Lambda_{QCD}\approx
(200 - 300)\,\mbox{MeV}$ and $m_b(m_b)\approx (4.5 - 5.3)\,
\mbox{GeV}$ \cite{PDG93}. We emphasize that this is a {\it
crude} numerical estimate for three reasons: First, the ambiguity
of the Borel sum or the minimal term of the series can only give
an indication of the size of the expected nonperturbative effects.
Second, the numerical
coefficient receives corrections of order $1/N_f^2$. Third, the
$1/N_f$ expansion does not provide us with the correct {\it nature}
of the IR renormalon singularity in general -- e.g., to
all orders in $1/N_f$ one expects the pole
to turn into a branch point. Therefore we do not control factors
of $\alpha(m_b)$ on the r.h.s. of
eq.({\ref{poleambiguity}), which can produce a substantial change.
Nevertheless,  the range of values quoted in eq.(\ref{poleambiguity}) should
give the right order-of-magnitude estimate for the ambiguities
inherent to the concept of the pole mass. The most important, but maybe
not too surprizing statement \cite{BIG94}, is that this ambiguity is of
order $\Lambda_{QCD}$ and not, say, $\Lambda^2_{QCD}/m$.

 There are a number of simple conclusions to be drawn from the presence
of the IR renormalons in the pole mass which still warrant a
discussion.
A matter of direct relevance is the calculation of total inclusive
widths of $B$-hadrons, which is receiving a lot of attention in the
literature.
Within perturbation theory the total decay widths are given simply by
the total widths for the free quark decay, expressions for which can be
taken over from QED studies of the muon decay

\be \label{semiwidths}
\Gamma(B\rightarrow X_q l \bar{\nu}_l) = \frac{G_F^2 |V_{bq}|^2
m^5_{b,\,pole}}{192\pi^3}\,(1+ \mbox{perturbative series})\,.
\ee

\n The result of principal interest, which triggered all later
discussions, is the observation \cite{BIG92} that the leading nonperturbative
effects in the
total widths are expressed in terms of the expectaion values of
dimension five operators of the kinetic energy and the chromomagnetic
interaction, and are down by two powers of the $b$-quark mass compared
to the perturbative contribution. Within the operator product
expansion there is no way to obtain corrections of order $1/m_b$, and
it is widely believed (see, e.g., \cite{MAN94}) that perturbation
theory is accurate to $1/m^2_b$ accuracy for the
total widths, provided the mass parameter which factors
eq.(\ref{semiwidths}) coincides with the pole $b$-quark mass.
Moreover, the
$1/m^2_b$ corrections prove  to be quite small.
In this situation it is appealing to try to determine the pole
mass from the experimental
data on the total widths, a task undertaken for instance in \cite{LUK94}.

The presence of the IR renormalon in the pole mass invalidates this
program. It implies that the ambiguity in the perturbative
series defining the pole mass inevitably induces an uncertainty of
order $\Lambda_{QCD}/m_b$ for the decay widths. The lesson which
should be learnt from the operator product expansion approach of
ref. \cite{BIG92} is that the {\em difference} in the total decay
widths of different $B$-hadrons is a $O(1/m_b^2)$ effect, while the
question of the absolute normalization is left open. In fact, one could
hope that the uncertainties in the summation of the perturbative
series in eq.(\ref{semiwidths}) compensates exactly the
uncertainties in the pole mass, rendering the perturbative prediction
unambiguous (up to $1/m_b^2$ accuracy) when expressed in terms of the
running renormalized quark mass at the scale $m_b$\footnote{
V.B. is grateful to N.G.Uraltsev for a discussion of this point. We
understand that a detailed study of this issue will be presented in the
work \cite{BIG94b}, and we gratefully acknowledge receiving a
preliminary version of this paper prior to its publication.}.
This question deserves further study.

Thus, the data on the total decay widths can not be used to {\em determine}
the pole $b$-quark mass, but rather can provide one with a one more
{\em definition} of it (using the truncated series in
eq.(\ref{semiwidths})). In
this respect, this definition is as good as any other phenomenological
definition, e.g.
coming from the studies of B-mesons or mesons of the $\Upsilon$ family in the
framework of the QCD sum rules \cite{VOL80,REI85,NAR86}.
The existing estimates for the $b$-quark mass span the range $m_b =
4.55-4.85$ GeV, and there has been
much debate on which of these values should be preferred.
In view of
eq.(\ref{poleambiguity}) a difference of $\delta m_{pole}\approx$
few hundred MeV can easily be attributed
to the ambiguity of the definition of a quantity, called ``pole mass'',
beyond perturbation theory. Thus, any claim for $m_{pole}$ with better
accuracy should be considered as hazardous, unless the precise
meaning of this quantity is specified.

\subsection{Matching and the residual mass term}

The self-energy of a massive quark contains powers of
logarithms of the type
$\ln(m^2-p^2)/m^2$ (times factors of $(m^2-p^2)/m^2$),
which are large, when the quark is heavy and near
mass shell, $p^2-m^2\approx m\Lambda_{QCD}$. HQET is designed to deal
with these large logarithms. To this end, one introduces a
factorization scale $\mu$ and writes

\be \ln \frac{p^2-m^2}{m^2} = \ln \frac{p^2-m^2}{m\mu} +
\ln \frac{\mu}{m}\,.
\ee

\n The first logarithm is small near mass shell, when $\mu\approx
\Lambda_{QCD}$ is taken and
the machinery of renormalization group techniques can then be
applied to sum the large logarithms of the type $\ln(m/\mu)$.
Remarkably, this
factorization can be achieved for the Borel transforms,
eq.(\ref{massiveselfenergies}), using an identity that relates
hypergeometric functions with argument $z$ and $1-z$. We obtain

\bea\label{factorizedselfenergies}
\tilde{\Sigma}_1(p^2,m,u) &=& \pref e^{-u C} \,3\,\Bigg\{
\left(\mmu\right)^{-u}\frac{\Gamma(u)\Gamma(1-2 u)}{\Gamma(2-u)}
\,{}_2 F_1\left(u,1+u,2 u;1-\frac{p^2}{m^2}\right) \nonumber\\
&&\hspace*{-2.5cm}
+ \,\left(\frac{m^2-p^2}{m^2}\right)
\left(\frac{m^2-p^2}{m\mu}
\right)^{-2 u}\frac{\Gamma(1-u)\Gamma(-1+2 u)}
{\Gamma(1+u)}\,{}_2 F_1\left(2-u,1-u,2-2 u;1-\frac{p^2}{m^2}\right)
\Bigg\}\nonumber\\
&&\hspace*{-2.5cm}
+\,\tilde{\Sigma}_2(p^2,m,u) + \pref\left(
-\frac{3}{u}+R_{\Sigma_1}(u)-R_{\Sigma_2}(u)\right)\\
\tilde{\Sigma}_2(p^2,m,u) &=& \pref e^{-u C}\,(-3 u)\,\Bigg\{
\left(\mmu\right)^{-u}\frac{\Gamma(u)\Gamma(1-2 u)}{\Gamma(3-u)}
\,{}_2 F_1\left(u,2+u,2 u;1-\frac{p^2}{m^2}\right) \nonumber\\
&&\hspace*{-2.5cm}
+ \,\left(\frac{m^2-p^2}{m^2}\right)
\left(\frac{m^2-p^2}{m\mu}
\right)^{-2 u}\frac{\Gamma(1-u)\Gamma(-1+2 u)}
{\Gamma(2+u)}\,{}_2 F_1\left(3-u,1-u,2-2 u;1-\frac{p^2}{m^2}\right)
\Bigg\}\nonumber\\
&&\hspace*{-2.5cm}
 + \,\pref\,R_{\Sigma_2}(u)\,.\nonumber
\eea

\n In the heavy quark limit $1-p^2/m^2\approx\Lambda_{QCD}/m$ (provided
$m$ is chosen judiciously, see below) and the series expansion
of the hypergeometric function realizes directly the heavy quark
expansion. At each order, expansion of the Borel transform in $u$
produces
two series containing  logarithms of either $m^2/\mu^2$ or
$(p^2-m^2)/(m\mu)$ only, thus completing the factorization to
all orders in the heavy quark expansion and to all
orders in perturbation theory in $a$ (but to leading order in $1/N_f$).
Before we can
construct the matching explicitly, we have to discuss the choice of
the expansion parameter.

In a heavy meson most of its momentum $p$ is carried by the heavy
quark, thus write $p=m_Q v+k$. Fixing the velocity $v$ of the heavy
quark (thereby selecting
a sector in the Hilbert space of the effective theory once and forever),
we are still left with some freedom to choose $m_Q$. We do not want
the residual momentum $k$ to scale with the heavy mass, so intuitively
we guess that $m_Q$ should be a ``physical" mass. In perturbation
theory, it does not matter, whether we take the pole mass or the
meson mass, but if we want to do better and account for terms of
order $\Lambda_{QCD}/m_Q$ consistently, a precise definition of $m_Q$
must be given. For the time being, we satisfy ourselves with
the observation, that after this is done, we could expand $m_Q$ in
a double series in the coupling and $\Lambda_{QCD}/m$, where $m$
is the renormalized mass, of the form

\be\label{mq}
m_Q = m\left(1+\sum_n s_n a(m)^{n+1}\right) - \delta m +
O\left(\frac{\Lambda_{QCD}^2}{m}\right)\,.
\ee

\n This fixes the parameter $m_Q$ once and forever and different
choices of $m_Q$ define different heavy mass expansions.
We call

\be\label{deltam}
\delta m = C m e^{1/(2\beta_0\alpha(m))}\alpha(m)^{b^\prime}
(1+O(\alpha)) = C\Lambda_{QCD}
\alpha(m)^b (1+O(\alpha))
\ee

\n the residual mass term and $C$, $b$ ($b^\prime$) are constants that
depend on
the definition of $m_Q$ (in particular, they could be zero)\footnote{
Taking $\delta m$ to be of order $\Lambda_{QCD}$ anticipates
eq.(\ref{poleequal}). Here it is only important to note that, if
$\delta m$ is non-zero, it is exponentially small in $\alpha$.}. With
these definitions at hand, we can continue and expand the self-energies,
given in eq.(\ref{factorizedselfenergies}), in $k/m_Q$. To be
precise, we will consider the inverse propagator, see eq.(\ref{fullprop}),
sandwiched between two projectors $(1+\!\not\!v)/2$, and define

\be \proj\,S^{-1}_P(v k,m_Q) = \proj\,S^{-1}(p,m)\,\proj\,.
\ee

\n After a little algebra, we arrive at the following expression
for the Borel transform of the inverse propagator:

\bea\label{main}
\tilde{S}^{-1}_P(v k,m_Q,u) &=& \tilde{m}_Q\left(\frac{m}{\mu},u\right)
\nonumber\\
&&\,\hspace*{-3.2cm} - \,m \left(\delta(u) + \pref\left[
\left(\frac{m_Q^2}{\mu^2}\right)^{-u} \!\!e^{-u C}\,6\,(1-u)\,\frac{
\Gamma(u)\Gamma(1-2 u)}{\Gamma(3-u)} - \frac{3}{u} +
R_{\Sigma_1}(u)\right]\right)\,\nonumber\\
&&\,\hspace*{-3.2cm} + \,v k\Bigg[ \delta(u) - \pref\,e^{-u C}\Bigg\{
\left(\frac{m_Q^2}{\mu^2}\right)^{-u} 6\,(-1+u^2)\,
\frac{\Gamma(u)\Gamma(1-2 u)}
{\Gamma(3-u)}\nonumber\\
&&\hspace*{0.5cm}+\left(-\frac{2 v k}{\mu}\right)^{-2 u} (-6)\,
\frac{\Gamma(1-u)\Gamma(-1+2 u)}{\Gamma(2+u)}\Bigg\}
- \pref \,R_{\Sigma_2}(u)\Bigg]\nonumber\\
&&\hspace*{-3.2cm}+\, O\left(\frac{(v k)^2}{m_Q},\frac{1}{N_f^2}\right)\\
&&\hspace*{-3.2cm}\equiv\,\tilde{m}_Q\left(\frac{m}{\mu},u\right) -
\tilde{m}_{pole}\left(\frac{m}{\mu},u\right)\,+\,\tilde{C}\left(
\frac{m_Q}{\mu},u\right)\star\tilde{S}_{eff}^{-1}(v k,u)\,+\,
O\left(\frac{(v k)^2}{m_Q},\frac{1}{N_f^2}\right) \nonumber
\eea

\n Here

\bea\label{c}
\tilde{C}\left(\frac{m_Q}{\mu},u\right)&=& \delta(u) - \pref\,e^{-u C}
\left(\mmu\right)^{-u} 6\,(-1+u^2)\,\frac{\Gamma(u)\Gamma(1-2 u)}
{\Gamma(3-u)}\nonumber\\
&&\,+ \pref\left(-\frac{3}{u} + R_{\Sigma_{eff}}(u) -
R_{\Sigma_2}(u)\right)\nonumber\\
\tilde{S}_{eff}^{-1}(v k,u)&=& v k\,\delta(u) - \tilde{\Sigma}_{eff}(v k,u)
\,,
\eea

\n and $\tilde{m}_{pole}$ and $\tilde{\Sigma}_{eff}$ have been defined
in eqs.(\ref{polemass}) and (\ref{heavyself}). The ``$\star$'' denotes the
convolution product\footnote{$\tilde{f}\star\tilde{g}$ is the Borel
transform of $f\cdot g$ and is given by
$(\tilde{f}\star\tilde{g})(u) =
\intl_0^u\dd u^\prime \tilde{f}(u^\prime)\cdot\tilde{g}(u-u^\prime)$
.}. Finally $\tilde{m}_Q$ stands for the Borel transform of the
series in eq.(\ref{mq}).
The residual mass term is exponentially small in the coupling and
therefore not seen in the ``perturbative'' definition of the Borel
transform, which we use throughout this paper. Therefore including possible
terms of order $\Lambda_{QCD}$ from the definition of $m_Q$, we recover the
inverse propagator through

\be\label{propinv}
S_P^{-1}(v k,m_Q,a) = \intl_0^\infty \dd t\,e^{-t/a}\,
\tilde{S}_P^{-1}(v k,m_Q,-\beta_0 t)\, - \,\delta m\,+\,
O\left(\frac{\Lambda^2}{m_Q}\right)\,.
\ee

\n Eq.(\ref{main}) is crucial for understanding the structure of
renormalon singularities in the heavy quark limit and is worth being
discussed in great detail.
Assume first that the expansion parameter $m_Q$ equals the
renormalized mass $m$, i.e. $\tilde{m}_Q=m\delta(u)$ and $\delta m=0$.
Let us list the following observations:

(1) In perturbation theory $u$ should be considered as infinitesimal and
factors like $(m_Q^2/\mu^2)^{-u}$ turn into a series in $\ln(m_Q^2/\mu^2)$,
when $\tilde{S}_P^{-1}$ is expanded in $u$ to yield the perturbative
expansion of $S_P^{-1}$ in $a$. Eq.(\ref{main}) has a very simple
structure: The first two lines scale with $m_Q$ and  are given by
$\tilde{m}_Q - \tilde{m}_{pole}$. The term proportional to $v k$ has
a factorized form and can be written as the product of a coefficient
function $\tilde{C}$, that contains all the (logarithmic) dependence
on $m_Q$, and the effective inverse propagator, which is
$m_Q$-independent. These terms appear as a sum and not as a product
in eq.(\ref{main}), because we neglect terms of order $1/N_f^2$,
cf. eq.(\ref{c}). It is evident from eq.(\ref{factorizedselfenergies})
that this factorization holds true in higher orders in the
$1/m_Q$-expansion, where to order $(v k)^2/m_Q$ it matches
onto the kinetic and magnetic energy contribution to the self-energy
of a heavy quark.

(2) The term porportional to $v k$ is finite at $u=0$ as it must be
for the renormalized inverse propagator. However, the two terms
in curly brackets -- corresponding to coefficient function and
effective propagator\footnote{In a slight abuse of language, we shall
refer to the effective quantities that depend on $v k$ also as
``matrix elements".} -- have poles at $u=0$ separately. Factorization
has introduced UV divergences into coefficient functions and
matrix elements. By subtracting and adding a term $(-3)/u+R_{\Sigma_{eff}}(u)$
to the expression in curly brackets as indicated in eq.(\ref{c}),
one chooses a particular factorization scheme. As known from many
other examples there is an arbitrariness in the separation of
contributions to coefficient functions and matrix elements, which
here is represented by the arbitrary function $R_{\Sigma_{eff}}(u)$.
In the language of HQET, a particular factorization scheme corresponds
to a particular wave function renormalization of the effective
heavy quark field $h_v$.

(3) Consider now eq.({\ref{main}) at finite $u$. In this way we
probe the {\it asymptotic} behaviour of the perturbative expansion
(in $u$ or, equivalently, in $a$) and explore the nonperturbative
effects which are seen by the renormalon singularities. In view of
our previous discussion, we are mainly interested in the point
$u=1/2$, but the effect of factorization on renormalons is quite
general: it introduces {\it new infrared} renormalons into the
coefficient functions (from $\Gamma(1-2 u)$) and {\it ultraviolet}
renormalons into the matrix elements (from $\Gamma(-1+2 u)$), which
are not present in the non-expanded inverse propagator. Thus these
singularities must cancel each other, just as the singularity at
$u=0$ does, and indeed they do, but the cancellation takes place
among {\it different} orders in the $1/m_Q$-expansion. For example,
the IR renormalon in the pole mass at $u=1/2$ cancels the UV
renormalon at $u=1/2$ in $\tilde{S}^{-1}_{eff}$ and similarly, the
singularity in $\tilde{C}$ cancels an UV renormalon in the matrix
elements in the next order in $1/m_Q$. At $u=3/2$ the cancellation
takes place over three orders in $1/m_Q$. Of course, the pole
at $u=1$ is not fully eliminated, since it is present in
$\tilde{S}^{-1}_P$
as mentioned previously.

(4) Factorization thus affects the structure of renormalons in a
very similar way as we are used to in finite orders of perturbation
theory (a small neighbourhood of $u=0$) with the difference that
the divergences are related over different orders in the expansion
parameter $m_Q$. It is tempting to introduce a factorization
scheme for renormalon poles in general, just as one usually does in
perturbation
theory, i.e. regarding the ``renormalon" pole at $u=0$ in particular.
For instance, the replacements

\bea
\tilde{m}_{pole}\left(\frac{m}{\mu},u\right) &\longrightarrow&
\tilde{m}_{pole}\left(\frac{m}{\mu},u\right)\,+\, \mu\,\pref
\frac{4}{1-2 u}\nonumber\\
\tilde{S}_{eff}^{-1}(v k,u)&\longrightarrow&\tilde{S}_{eff}^{-1}(v k,u)
\,+\, \mu\,\pref\frac{4}{1-2 u}
\eea

\n will eliminate the divergence coming from $u=1/2$
from the perturbative expansions of the pole mass and the effective self-energy
without affecting $\tilde{S}_P^{-1}$. Technically, this can be
achieved within dimensional regularization by subtraction of
the poles at $d=3$. However,
such a substitution messes up the $1/m_Q$-expansion, since
it introduces {\it powers} of the scale $\mu$. In fact, it
acts analogously to a hard cutoff in the Feynman integrals, which
removes the contribution to the pole mass from the IR region and
from the UV region to the self-energy. We recognize that the present
discussion of the heavy mass expansion
parallels the discussion of the short-distance expansion in Sect. 2 and
indeed the renormalons appear for the very same reason, that, e.g.
for the coefficient functions, diagrams with a large number of bubbles
are dominated by internal momenta smaller than $\mu$. Quantitatively,
the crucial difference to the SDE is that the infrared effects
appear at order $\Lambda_{QCD}/m_Q$ (take $\mu=m_Q$) and not at order
$\Lambda_{QCD}^4/Q^4$.

After these remarks on the heavy mass expansion in general, we are in the
position to consider the construction of HQET. The effective
Lagrangian, eq.(1.1), is independent of $m_Q$ and supposed to extract
the correct dependence of the Green functions on $m_Q$ to leading
order. From eq.(\ref{main}) it follows that this purpose cannot be
accomplished with the
renormalized mass as the expansion parameter, but one {\it must} choose
$m_Q$ such that it
coincides with the pole mass to all orders of perturbation theory,

\be \label{poleequal}
\tilde{m}_Q\left(\frac{m}{\mu},u\right) =
\tilde{m}_{pole}\left(\frac{m}{\mu},u\right)\,,
\ee

\n in order to cancel the term that scales
with $m_Q$. This results in\footnote{It is sometimes understood that
the coefficient function for the
propagator is unity. This can be achieved by a
particular choice of the
renormalization scheme in QCD {\it via} the function $R_{\Sigma_2}(u)$,
see eq.(\ref{main}). Clearly, this requires a mass-dependent
scheme with non-analytic counterterms (see App. A). and for this
reason we refrain from performing this step, when we discuss the
singularities of the Borel transform. It is not strictly necessary
and will not be important for the following. }

\be\label{invpropsub}
\tilde{S}_P^{-1}(v k,m_Q,u) = \tilde{C}\left(
\frac{m_Q}{\mu},u\right)\star\tilde{S}_{eff}^{-1}(v k,u)\,+\,
O\left(\frac{(v k)^2}{m_Q},\frac{1}{N_f^2}\right) \,.
\ee

\n The price to pay for the elimination of all terms scaling with $m_Q$
in perturbation theory is, however, that
 the renormalon singularity at $u=1/2$ is no longer
cancelled. With $\tilde{S}_P^{-1}$ of eq.(\ref{invpropsub}) inserted
into eq.(\ref{propinv}), the Borel integral is no longer well-defined
up to $u=1$ as it was with the non-expanded inverse propagator
(or the choice $m_Q=m$). We have introduced a spurious renormalon
into the construction of HQET, which renders
the Borel integral ambiguous by
terms of order $\Lambda_{QCD}$! This is not a big surprize, because
the Borel integral for $m_Q$ is itself ambiguous by terms of this order,
see eq.(\ref{poleequal}), and the discussion of the previous subsection.
It is now high time to return to the definition of $m_Q$, eq.(\ref{mq}).

Once we want to use HQET beyond perturbation theory and include
corrections of order $\Lambda_{QCD}/m_Q$ an unambiguous definition
of $m_Q$ becomes imperative. Let us therefore imagine a summation
prescription for the divergent expansion in eq.(\ref{mq}) that
{\it defines} $m_Q$ as a function of $a$ with certain analyticity
properties. This fixes the exponentially small in $a$ terms, which
are not seen in the perturbative expansion. We can accomplish this
with Borel summation, if we add a residual mass term $\delta m$ of
order $\Lambda_{QCD}$
as in
eq.(\ref{mq}) and understand that it is formally ambiguous and
possibly complex. To explain this, recall that one must
give a prescription for the pole in the Borel integral at $u=1/2$.
The Borel integral will differ whether one chooses to deform the
contour into the upper or lower complex plane and the residual mass
term must cancel this ambiguity, if $m_Q$ itself is unambiguous\footnote{
This tacitly assumes that an asymptotic expansion of $m_Q$ in $a$ can be
performed in a sector around the positive real $a$-axis and that the
fixed sign divergence for positive $a$ is indeed correlated with
a Stokes discontinuity in the exponentially suppressed terms. We are
far from proving such a statement, which is, however, an underlying
assumption in practically all works, which relate the asymptotic behaviour
of perturbation series to ``nonperturbative terms'' and
single out the Borel summation and its extensions to series
with fixed sign divergence. It has been
verified in model calculations \cite{DAV84}, that this gives the
correct prescription, but is conjecture beyond.}. This residual
mass term is {\it obligatory} and its presence in eq.(\ref{propinv})
cancels precisely the ambiguity from the spurious pole at $u=1/2$
in the Borel sum. In this sense the status of the residual mass term
is identical to the status of the condensates in the SDE. Recall
that the ambiguity in the definition of condensates compensates
the IR renormalons in the perturbative expansion. It is important
that $\delta m$ is not a physical quantity. It is defined as a
number only after one has fixed a summation prescription for the
perturbative expansion of the Green functions in HQET (which, of
course, should be done consistently). Apart from this formal analogy,
the residual mass term is very different from condensates in the SDE
and certainly can not be related to condensates in any way. It has
no {\it direct} dynamical origin, but it is through this ambiguous
residual mass, that HQET remembers that the concept of a ``quark on
mass-shell'' is not physical, even if the quark is very heavy. The
modifications of the effective Lagrangian of HQET in the
presence of a residual mass term have already been given in
ref.\cite{FAL92}. If we understand the ambiguous nature of $\delta m$,
we can in fact copy all expressions given there. In particular,
the leading order Lagrangian, eq.(\ref{effectivelagrangian}), has to
be modified in the obvious way

\be \label{neweffectivelagrangian}
{\cal L}_{eff}\,=\,\bar{h}_v i v\cdot D h_v - \delta m \bar{h}_v h_v
+ {\cal L}_{light}
\,.\ee

\n The appearance of an ambiguous quantity in the Lagrangian might seem
peculiar. But this Lagrangian arises as a {\it remnant} of QCD. Without
fixing a specific summation prescription for the Green functions
in QCD {\it before} the construction of HQET, the ambiguous residual
mass appears in the above Lagrangian to render the Green functions
derived from this Lagrangian {\it unambiguous} and invariant under
different summation prescriptions.

To recognize the implications of an ambiguous residual mass
term, define $\bar{\Lambda}$ as the difference of the heavy hadron
and the heavy quark expansion parameter in the heavy mass limit:

\be\label{lamb}
m_P - m_Q = \bar{\Lambda} + O\left(\frac{1}{m_Q}\right)
\ee

\n By construction, $\bar{\Lambda}$ is well-defined, but it inherits
the arbitrariness inherent in the definition of $m_Q$. It has been
shown \cite{FAL92} that the matrix elements of HQET depend only on the
combination

\be \bar{\lambda} = \bar{\Lambda} - \delta m \,,\ee

\n which is invariant under redefinitions of $m_Q$. It
turns out to be an important quantity for the parametrization
of the matrix elements to order $1/m_Q$. For example, it is the
only unknown parameter governing the $1/m_Q$-corrections to the
decays of a heavy baryon, where the light quarks are in a spin zero
state \cite{GEO90b}. We conclude from our above analysis that this
quantity is theoretically ambiguous by terms of order $\Lambda_{QCD}$
and has no physical meaning by itself. The ``operational" definition,
given in \cite{FAL92},

\be \bar{\Lambda} - \delta m\,=\,\frac{\langle 0|\bar{q}(i v\cdot
\stackrel{\leftarrow}{D})
\Gamma h_v|M(v)\rangle}{\langle 0|\bar{q}
\Gamma h_v|M(v)\rangle},
\ee

\n where $|M(v)\rangle$ denotes a meson state, $q$ a light quark field and
$\Gamma$ a Dirac matrix, is, in fact, illusive. The matrix element
is ambiguous. It can not be directly related to any physical quantity and
defining it requires a nonperturbative regularization,
which can not avoid the renormalon problem.
If $\bar{\lambda}$ were a physical quantity and
could be unambiguously determined, this would provide us with a {\it unique}
nonperturbative
definition of a heavy quark mass, $m_{pole}$,
through eq.(\ref{lamb}) as is indeed widely
maintained (see e.g. \cite{MAN94,NEU93}). Unfortunately, this is not
so. The decay of a $\Lambda_b$-baryon \cite{GEO90b} may serve as an
illustration.
The value of $\bar{\lambda}$ enters the form factors of this decay
at subleading order in $1/m_Q$. The leading order form factors receive
perturbative corrections with a divergent series expansion which is
expected to have
a renormalon at $u=1/2$. Theoretically, corrections of order $\Lambda_{QCD}/
m_Q$ are well-defined only after the choice of a summation prescription
for this series. Unless this is done, $\bar{\lambda}$ must be ambiguous.
{\it Practically}, it might still be useful to fit a value for
$\bar{\lambda}$ phenomenologically where, again, we appeal to the analogy
with the condensates in the SDE. But there can be no rigorous
theoretical determination of $\bar{\lambda}$, as has already been
observed on physical grounds in ref. \cite{BIG94}, which in fact has
triggered our more formal
investigation.



\mysection{The correlation function of heavy-light currents in HQET}
{The correlation function of heavy-light currents in HQET}

While the quark propagator has been very useful to gain some
insight into the structure of the heavy quark expansion in large
orders of perturbation theory, it is not a quantity of particular
physical interest. We devote this section to the study of the
Borel transform of the correlation function of heavy-light currents
in HQET. Its spectral density contains a heavy meson pole and
the short-distance expansion of the correlation function provides
access to $\bar{\Lambda}$ and the decay constant of the meson
through the technique of the QCD sum rules. To be definite, we consider
the perturbative expansion to order $1/N_f$ of

\be\label{CF2}
\Pi_5(\omega)\,=\,i\int\dd^4 x\,e^{i\omega (v\cdot x)}\,
\langle 0|T\{j_5^\dagger(x) j_5(0)\}|0\rangle\qquad
j_5(x) = \bar{h}_v(x) i\gamma_5 q(x)\,.
\ee

\n The choice of the Dirac matrix turns out to be unimportant, since
the pseudoscalar and vector mesons are degenerate to leading order in
the $1/m_Q$-expansion. The variable
$\omega=v q$ has the meaning of a frequency, and
measures the off-shellness of the
heavy quark, provided that $m_Q$ has been unambiguously
defined as explained in the previous section and $q$ is the residual
momentum. We will not construct the full matching to QCD as for the
inverse propagator, which would require the calculation of the
corresponding correlation function in QCD with a massive quark to
obtain the coefficient function. $\Pi_5(\omega)$ is quadratically
divergent and the current product needs an additional subtraction,
which is a second-order polynomial in $\omega$. This subtraction can
be avoided by taking three derivatives and we shall consider

\be D(\omega)\equiv \omega\,\frac{\dd^3\Pi_5(\omega)}{\dd\omega^3}
\ee

\n in the following.

The SDE of the corresponding correlation function in QCD can be
repeated in HQET in a two-step procedure. First, the momenta larger
than $m_Q$ are integrated out, which results in a series in $1/m_Q$ of
correlation functions of operators in the HQET of which $\Pi_5(\omega)$
is the first term. Second, the products of effective operators
are expanded at short distances, that is $\Lambda_{QCD}\ll\omega < m_Q$.
The SDE of $\Pi_5(\omega)$  is given by  \cite{BAG92}

\bea\label{opehl}
D\left(\frac{\omega}{\mu},\alpha(\mu)\right)\!\!&=& \!\!-\frac{3}{\pi^2}
\left(1 + \frac{\alpha(\mu)}{\pi}\left\{\frac{8}{3} + \frac{4\pi^2}{9}
- 2\ln\left(-\frac{2\omega}{\mu}\right)\right\} + O(\alpha^2)\right)\\
&&\!\!-\,\frac{3}{\omega^3}\,\langle\bar{q} q\rangle(\mu)
\left(1 + 2\frac{\alpha(\mu)}{\pi} + O(\alpha^2)\right) +
\frac{15}{4\omega^5}\,\langle g\bar{q}\sigma G q\rangle(\mu)
\left(1 + O(\alpha)\right) + \ldots \,,\nonumber
\eea

\n where the omitted series of higher dimensional operators starts
with four-quark operators. Since the heavy-light current acquires
an anomalous dimension in the effective heavy quark theory, the
two-loop perturbative correction is now scheme-dependent in contrast
to the case of vector currents of light quarks, and the above result
is given in the $\overline{MS}$-scheme. Note also that to leading
order in the $1/m_Q$-expansion there is no contribution from the
gluon condensate.

As familiar by now, to order $1/N_f$ we are interested in the contribution
from dia\-grams with an arbitrary number of light quark loops inserted
into the gluon line of the two-loop diagrams. The Borel transform
of this class of diagrams can conveniently be computed by inserting
the Borel-transformed gluon propagator, eq.(\ref{gluonprop}), see
Fig. 4. The remainder is technical. On the one hand the calculation
is far less tedious than in the light-quark case \cite{BEN93c},
since the spinor structure simplifies in the heavy quark limit. On
the other hand, the correlation function has an anomalous dimension and
one looses the Ward identity (which holds in QCD to order $1/N_f$),
which ensured that all divergences cancel after one subtraction of
the correlation function of light quark currents. The correlation
function of the effective heavy-light currents needs an explicit
renormalization and we refer again to App. A, where the procedure is
outlined. It turns out that only the diagram (c) in Fig. 4 has a pole
at $u=0$, which is eliminated in this way. The most difficult part
comes from the non-reducible scalar part of the diagram (a). The details
of its computation are given in App. B.

The result for the Borel transform of the correlation function is

\be
\tilde{D}\left(\frac{\omega}{\mu},u\right) = -\frac{N_c}{\pi}\,
\delta(u) + \tilde{D}_{(a)}\left(\frac{\omega}{\mu},u\right) +
\tilde{D}_{(b)}\left(\frac{\omega}{\mu},u\right) +
\tilde{D}_{(c)}\left(\frac{\omega}{\mu},u\right) + O\left(
\frac{1}{N_f^2}\right)
\ee

\n with the separate contributions (in the Landau gauge) from the three
diagrams shown in Fig. 4 given by ($N_c=3$, $C_F=4/3$)

\bea
\tilde{D}_{(a)}\left(\frac{\omega}{\mu},u\right)&=&\frac{C_F N_c}
{4\pi^3 N_f}
\Bigg[\omu e^{-u C}\Bigg\{2 u (1-2 u) (2-2 u)\left[S(4,u)-S(4,1+u)
\right]\nonumber\\
&&\hspace*{1.5cm} +\,\frac{\Gamma(-u)\Gamma(1+2 u)}{\Gamma(2+u)}
\left[3+5 u-\frac{1+u}{1-2 u}\right]\Bigg\} + R_{(a)}(u)\Bigg]
\nonumber\\
\tilde{D}_{(b)}\left(\frac{\omega}{\mu},u\right)&=&\frac{C_F N_c}
{4\pi^3 N_f}
\Bigg[\omu e^{-u C}\,3\,\frac{\Gamma(1-u)\Gamma(1+2 u)}{(2-u)
\Gamma(2+u)} + R_{(b)}(u)\Bigg]\\
\tilde{D}_{(c)}\left(\frac{\omega}{\mu},u\right)&=&\frac{C_F N_c}
{4\pi^3 N_f}
\Bigg[\omu e^{-u C}\,6\,\frac{\Gamma(1-u)\Gamma(-1+2 u)}{
\Gamma(2+u)} + \frac{3}{u} + R_{(c)}(u)\Bigg]\,,\nonumber
\eea

\n where

\bea
S(4,\gamma) &=& -\Gamma(1-\gamma)\Gamma(2-\gamma)\Gamma(\gamma-1)^2\\
&& + 2\,\frac{\Gamma(1-\gamma)\Gamma(2\gamma-1)}{\Gamma(\gamma-1)}
\sum_{n=0}^\infty \frac{1}{\gamma-1+n}\,\frac{\Gamma(2\gamma-2+n)}
{\Gamma(3\gamma-2+n)}\,\frac{\Gamma(\gamma-1+n)}{n!}\,,\nonumber
\eea

\n (see eq.(\ref{scalarintegral})). The $R$-functions are scheme-dependent
and arbitrary in general apart from being non-singular in a neighbourhood
of $u=0$. In $MS$-like schemes, they are entire and their values at
$u=0$ can be found to be $R_{(a)}(0)=0$, $R_{(b)}(0)=-3/2$,
$R_{(c)}(0)=4$. A check of our result is provided by the value of
the Borel transform at $u=0$ (disregarding the $\delta$-function),
which must reproduce the two-loop perturbative correction to $D$.
With the help of the expansions collected in App. B, we find

\bea
\tilde{D}_{(a)}\left(\frac{\omega}{\mu},0\right)\!&=&\!\frac{3 C_F N_c}
{4\pi^3 N_f}\left\{\frac{4}{3} + \frac{4\pi^2}{9}\right\}
\qquad \tilde{D}_{(b)}\left(\frac{\omega}{\mu},0\right)\,=\,0
\nonumber\\
\tilde{D}_{(c)}\left(\frac{\omega}{\mu},0\right)&=&\frac{3 C_F N_c}
{4\pi^3 N_f}\left\{-\frac{1}{3} - C - 2\ln\left(-\frac{2\omega}{\mu}
\right)\right\}\,,
\eea

\n in agreement with eq.(\ref{opehl}) in the $\overline{MS}$-scheme
($C=-5/3$).

We now turn to the discussion of the renormalon singularities in the
Borel plane. A summary of this discussion is presented in Fig. 5,
where the Borel plane for the correlation function of heavy-light
currents in HQET is compared with the situation for light quarks in QCD.

{\it Infrared renormalons.} The IR singularities occur at positive
integers starting from $u=3$ and, generically, are double poles.
The poles at $u=1$ and $u=2$ are present in every single diagram but
cancel in the sum of all three. In general, a condensate of
dimension $d$ in the SDE can be related to an IR renormalon at
$u=d/2$. By comparison with eq.(\ref{opehl}) we find that the
quark condensate and the mixed quark-gluon condensate do not
produce IR poles in the coefficient function of the unit operator.
This is physically clear, because the renormalons originate from
a soft gluon line in the diagrams of Fig. 4. More formally, the
renormalons are linked to an ambiguity in the definition of the
vacuum expectation values of composite operators due to mixing
with lower dimensional operators in the sense that the definition
of condensates in principle requires a prescription for the sum
of all perturbative series that appear in lower dimensional terms.
The operator $\bar{q} q$ can not mix with any lower dimensional
operator due to its different transformation properties under
chiral symmetry\footnote{Complications are bound to arise, if the
regularization breaks chiral symmetry.} and its vacuum expectation
value is unambiguously determined by the pion decay constant
\cite{DAV84} through the PCAC relation. The mixed quark-gluon operator
$g\bar{q}\sigma G q$ has no such protection, but due to its chiral
transformation properties mixes only with $\bar{q} q$. For this reason,
the mixed quark gluon condensate is not seen as an IR renormalon
in the perturbative expansion, but should be related to an IR
renormalon in the Borel transform of the coefficient function of
the chiral condensate at $u=1$.

The cancellation of the IR renormalon at $u=2$ can be directly
attributed to the absence of the gluon condensate in the SDE to
leading order of the $1/m_Q$-expansion. Thus, all IR renormalons
in the correlation function are in complete agreement with the SDE.
The first singularity at $u=3$ comes from four-quark operators
not written in eq.(\ref{opehl}). Since there are no IR renormalons
at half integers, we conclude that odd dimensional operators do
not produce IR poles in the coefficient function of the unit
operator, which is again a consequence of chiral symmetry. If we
assume that the equation of motion has been used to reduce all
operators with covariant derivatives, then odd dimensional gauge-invariant
operators must contain $4 k+2$ ($k=0,1\ldots$) quark fields and
an arbitrary number of gluon field strenghts and a chiral-invariant
operator with this number of quark fields can not be constructed.

{\it Ultraviolet renormalons.} In contrast to the light quark case,
UV renormalons appear in the perturbative expansions in HQET at
all negative half integers for the simple reason that there is a
dimension one parameter $\omega$ available. In addition, one finds
a non-Borel-summable UV renormalon at $u=1/2$ on the positive axis,
which, in fact, stems only from the diagram (c). The pole at $u=1/2$
in some terms of $\tilde{D}_{(a)}$ is spurious and drops out in the
full expression. This UV renormalon is a simple pole and can be traced
to the insertion of the self-energy of the heavy quark in the
diagram (c), which we have investigated in detail in Sect. 3, where
the linear divergence of the self-energy has been identified as the
cause of this pole. Note that the correlation function $\Pi_5(\omega)$
is quadratically divergent and has UV poles at $u=1/2$ and $u=1$, which
have been eliminated by taking three derivatives. In a similar way
the UV renormalon at $u=1/2$ is removed from the first derivative
$\partial \tilde{\Sigma}_{eff}(v k)/\partial(v k)$ of the self-energy,
but since $\Sigma_{eff}(v k)$ and not its derivative is inserted
into the Green functions of HQET, there is no way to avoid the
UV renormalon generated by the linear divergence of the self-energy
to pervade to all Green functions in HQET.

The emergence of non-summable singularities on the positive Borel
axis signals that perturbation theory is incomplete and points
towards either inconsistency of the theory or some nonperturbative
phenomenon  which in a formal language cures the ambiguities of the
Borel integral. For the IR renormalons in HQET this is provided
by the condensates just as in QCD and they account for the nonperturbative
terms that arise in the SDE of the correlation functions. The
UV renormalon at $u=1/2$ reminds us of a nonperturbative effect
of a very different nature, which can not be attributed to
short distances: HQET (without a residual mass term) is an effective
theory for a heavy on-shell fermion that does not exist in nature
and the binding energy of a quark in a meson is not a physical concept.
Indeed, we have seen in Sect. 3 that one must add a residual mass
term $\delta m$ to the effective Lagrangian, see
eq.(\ref{neweffectivelagrangian}),
which we have omitted in our discussion so far. Since in the
$1/N_f$ expansion $\delta m$ counts as $1/N_f$, to order $1/N_f$ the
term

\be D_{\delta m}\left(\frac{\omega}{\mu},\alpha(\mu)\right) =
\frac{N_c}{\pi^2}\,\frac{\delta m}{\omega}
\ee

\n has to be added to eq.(\ref{opehl}), which cancels the ambiguity
of the Borel integral for $D$ due to the UV renormalon.
We repeat that the residual
mass term is formally of order $\Lambda_{QCD}$ and serves two purposes:
(1) It guarantees that the predictions of HQET are invariant under the
choice of the expansion parameter $m_Q$ \cite{FAL92}, (2) it ensures that the
predictions of HQET are invariant under the choice of summation
prescription for the UV renormalon divergence of the correlation
functions in HQET. In this respect, it acts analogously as the condensate
terms in eq.(\ref{opehl}) with respect to the IR renormalons. To be
precise, one could agree that all series should be summed by a contour
of the Borel integral through the upper complex plane, which would
fix $\delta m$ with an imaginary part that ensures reality of the
correlation functions and defines a particular $m_Q$. One can
convince oneself that if this is accomplished for the self-energy,
then it is automatic for $D$, where it is an important consistency
check that the diagrams (a) and (b) do not produce a pole at $u=1/2$.
Superficially the presence of a residual mass leads to new terms
of order $\Lambda_{QCD}/\omega$ in eq.(\ref{opehl}), which can not be
avoided because of the divergence of perturbation series and which
spoil the SDE, where it is assumed that all power corrections can be
accounted for by condensates. We shall show below that the power
corrections due to the residual mass are organized in a very particular
way and can effectively be summed up.

Let us first throw a glance beyond the $1/N_f$-expansion. Consider the
class of diagrams, where a second heavy quark self-energy is inserted into
the heavy quark line of the diagram (c) of Fig. 4 and with an
arbitrary number of fermion loops in any of the two gluon lines.
Apart from factors these diagrams can be obtained by squaring the
series in $\alpha$, implied by the diagram (c). The Borel transform is
given by the convolution of diagram (c) with itself and develops an
UV renormalon at $u=1/2$ and $u=1$. Obviously, this process can be
iterated and the diagrams with a chain of $n$ self-energies produce
a Borel transform with UV poles at all half integers up to $n/2$,
which are of course related to $n$ insertions of the residual mass
term. We conclude that to all orders in perturbation theory, the
UV renormalons proliferate and spread over all half-integers on the
negative {\it and} positive Borel axis. Opposite to the situation
with light quark currents, where the leading order in $1/N_f$
gives a complete picture of the renormalons in the Borel plane (as
far as we know), see Fig. 5a, the Borel plane of the correlation
function of heavy-light currents in HQET becomes modified to all
orders in $1/N_f$. As in QED, there is a series of UV renormalons
on the positive axis, but it must be emphasized that their physical
origin is so completely different that the common name is hard to
justify: In QED, the UV renormalons arise from the logarithmic increase
of the effective coupling in the UV region; in HQET all UV renormalons
are generated by the linear divergence of the heavy quark self-energy
and there is no relation to the effective coupling at all.

The effect of a residual mass term on the correlation function has an
almost trivial structure. To see this, let us for a moment ignore all the
complexities of the residual mass term and treat it as a number.
Multiple insertions of $\delta m$ into a heavy quark line can be
summarized by implementing $\delta m$ on the Lagrangian level as
already done in eq.(\ref{neweffectivelagrangian}), which modifies
the heavy quark propagator to

\be
\proj\,\frac{i}{v k-\delta m}\,.
\ee

\n Call $\Pi_5(\omega)$ the correlation function, computed from the
Lagrangian, eq.(\ref{effectivelagrangian}), without a residual mass
term and $\Pi_5^{\delta m}(\omega)$ the same object, computed from the
Lagrangian in eq.(\ref{neweffectivelagrangian}). Then

\be\label{scale}
\Pi_5^{\delta m}(\omega) = \Pi_5(\omega - \delta m)\,,
\ee

\n i.e. the sole effect of a residual mass of the effective heavy quark
is to produce a shift of the momentum scale in the correlation
function. The validity of eq.(\ref{scale}) is obvious on physical
grounds. Recall that $\omega$ is ``measured'' from the point $m_Q$
(if $p$ is the physical mometum of the meson, $v p = m_Q+\omega$). Thus
different choices of $m_Q$, leading to different values of $\delta m$,
simply shift the ``zero point'' of the momentum scale. If the predictions
of HQET are to be invariant under the choice of $m_Q$, this can only
result in a change of the argument of the correlation function.
Nevertheless a diagramatic proof of eq.(\ref{scale}) might be useful.
Let $q$ denote the residual momentum of the heavy quark, let
$\omega=v q$ and consider an arbitrary diagram $\Gamma$ that contributes
to the perturbative expansion of $\Pi_5^{\delta m}(\omega)$. Since
all diagrams with heavy quark loops vanish identically, the only
way the heavy quark can appear in $\Gamma$ is as a line that joins the
two current insertions and emitting an arbitrary number $m-1$ of gluons.
Now label the independent loop momenta of $\Gamma$ such that the
heavy quark propagators carry momentum $k_i+q$, $i=1,\ldots,m$ and
call $p_j$ the remaining loop momenta. With this assignment all other
propagators are independent of $q$ and the diagram can be represented
as

\be
\Gamma = \int\prod_{i=1}^m\dd k_i\prod_j\dd p_j\,\left(\prod_{i=1}^m
\frac{1}{v k_i+\omega - \delta m}\right)\times \tilde{\Gamma}(k_i,p_j)
\,,\ee

\n where the remaining part $\tilde{\Gamma}$ of the diagram is
independent of $\omega$ and $\delta m$. Therefore any diagram
depends only on the combination $\omega-\delta m$, proving
eq.(\ref{scale}).

Formally, the residual mass term is an ambiguous quantity and the
terms proportional to $(\delta m/\omega)^n\sim(\Lambda_{QCD}/\omega)^n$
are present in eq.(\ref{opehl})
to render $\Pi_5^{\delta m}(\omega)$ unambiguous and
well-defined. In practice, summation of perturbative expansions is
never performed, since only a few low order terms of the series
are available. Eq.(\ref{scale}) tells us that neglecting the UV renormalon
divergences in the perturbative series can be equally interpreted
as an uncertainty of order $\Lambda_{QCD}$ in the momentum scale
$\omega$ of the correlation function. In this sense, we say that the
UV renormalons can be ``summed up'' to produce an ambiguity of
scale in the HQET. Indeed, this scale ambiguity captures most concisely
the physics reflected in these UV renormalons. Finally, their
appearance can be traced back to the attempt to split a nonperturbative
residual momentum $k$ of order $\Lambda_{QCD}$ from the meson
momentum, attributing the remainder to the momentum of a ``physical''
quark. This is not an infrared safe procedure, as is clearly visible
from the IR renormalons in the pole mass.

At this point, comparison with the lattice formulation of HQET may help
to clarify the meaning of an ambiguous residual mass. The discretized
version of the heavy quark propagator has a linear divergence in the
lattice spacing starting from first order in perturbation theory
\cite{MAI92}, which comes from the presence of a dimensionful cutoff and
can be absorbed into a mass renormalization. The asymptotic behaviour
of correlation functions at large times is proportional to
$\exp(-\bar{\lambda} t)$, where $\bar{\lambda}$ is the mass of the
lowest excitation of the theory (see eq.(\ref{lamb})). The presence of
a linear divergence leads to the conclusion \cite{MAI92} that the
exponent $\bar{\lambda}$ is not a physical quantity.
It is evident that the UV renormalons, the ambiguous residual mass term
(and consequently $\bar{\lambda}$, see Sect. 3.3) and the scale ambiguity
of correlation functions are in fact a reflection in the continuum
of one and the
same phenomenon that has been observed a long time ago
on the lattice.



\mysection{The status of QCD sum rules}{The status of QCD sum rules}

The SDE of correlation functions has become an important tool to
determine the various nonperturbative parameters of HQET through
the QCD sum rules method. In the light of our
results some of the common lore about  the QCD sum rules,
when applied to HQET, needs to be
revised.
It is instructive to trace
uncertainties induced by the presence of UV renormalons in the SDE of
correlation functions in HQET on some particular sum rules. We start
with the simplest one, which is for the $B$-meson
coupling in the static limit.
An important result obtained from this sum rule is an
estimate of the quantity $\bar\Lambda $ (or $\bar{\lambda}$),
a task which has never
been accomplished on the lattice for the reasons mentioned above.

In the QCD sum rule approach one takes the Borel
transform of the SDE of the correlation function with respect to
$1/\omega$ (not to be confused with the Borel transform with respect
to the coupling, which has been discussed in the previous sections),
trading the frequency $\omega$ for a new variable,
the Borel parameter $\tau$.
This ``theoretical'' expression is matched to the
``phenomenological'' part of the sum rule, which uses a
dispersive representation of the correlation function and
saturation of the imaginary part by hadronic states. The
Borel transformation serves several purposes, and ensures that both
higher condensate contributions to the SDE and higher-mass contributions to
the expansion of the imaginary part in hadron states are suppressed.
The matching is performed in
a certain intermediate region of the Borel parameter, where one hopes
that both the SDE and the hadron expansion work reasonably well.
The effect of the UV renormalons on the sum rule for the
$B$-meson coupling (see, e.g.,
\cite{BAG92,NEU92}) can easily be seen by the use of eq.(\ref{scale}).
Indeed, since by virtue of the Borel transformation

\be
\frac{1}{(\delta m -\omega)^n}
 \rightarrow e^{-\delta m/\tau}\frac{1}{(n-1)!\,\tau^n}
\,,
\ee

\n the effect of the ambiguity in the scale $\omega $ transforms to an
overall factor $\exp(-\delta m/\tau)$ in front of the ``theoretical''
side of the sum rule. Thus, to be concrete, the sum rule for
the correlation function $\Pi_5$ in eq.(\ref{CF2}) is modified to

\bea\label{SRFB}
\widehat f^2_B e^{-\bar\Lambda/\tau}& =&
e^{-\delta m/\tau}
\Bigg\{\frac{3}{\pi^2}\int_0^{\omega_0}\!d\omega\,\omega^2
e^{-\omega/\tau}[1+\mbox{perturbative series}]\nonumber\\&&\mbox{}
 -\langle\bar q q\rangle(\mu=2\tau)
+\frac{1}{16\tau^2}\langle g\bar q \sigma Gq\rangle(\mu=2\tau)+
\ldots \Bigg\}\,,
\eea

\n where $\widehat f_B $ is the $B$-meson leptonic decay constant in the
static limit (at a low scale $\mu=2\tau $),
 and $\omega_0$ is the duality interval for the lowest
bound state. For simplicity, we have discarded the radiative
corrections, see \cite{BAG92} for the complete expression to two-loop
accuracy.

The factor $\exp(-\delta m/\tau)$ can be brought to the l.h.s. and combined
with $\exp(-\bar\Lambda/\tau)$ so that the sum rule depends on the
combination $\bar\Lambda-\delta m$ only, as expected. Since this
parameter is extracted from the sum rule by a fitting
procedure, one may conjecture that the effects of renormalons are
completely eliminated. As stressed repeatedly above, this conclusion
is wrong. The presence of UV renormalons in the perturbative series on
the r.h.s. of eq.(\ref{SRFB}) indicates a principal ambiguity in its
summation, which is expressed in a shorthand form by the appearance
of the ambiguous residual mass term. In other words, if one
calculated the corrections to the r.h.s. of the sum rule,
eq.(\ref{SRFB}), from large orders in the perturbative expansion (using some
prescription to sum the series, see above,
which also fixes $\delta m$ to a definite value), the main effect
of these corrections will be a change of the output value of
$\bar\Lambda-\delta m$ by an amount of order $\Lambda_{QCD}$.
Note that the coupling $\widehat f_B$ formally is protected from such
corrections -- the residue of the pole in the correlation function
does not depend (formally) on the position of this pole. In practice,
however, the values of $\bar\Lambda-\delta m$ and $\widehat f_B$
extracted from the sum rule are strongly correlated (see e.g. the
discussion in \cite{BAG92}), and an uncertainty of 100 MeV in
$\bar\Lambda-\delta m$ induces an uncertainty of order 15\% for the
static decay constant.

The observation that the effect of the
non-Borel-summable UV renormalons in HQET can
generally be ascribed to an ambiguous residual mass term, allows for a
back-on-the-envelope estimate of their importance in other sum rules,
which have a more complicated structure. As an example, let us consider
the sum rule for the heavy quark kinetic energy, which is defined by the
expectation value of the operator of the nonrelativistic kinetic energy in
the meson state. In the
presence of a residual mass term it is given by

\be\label{kinetic}
 K_{\delta m} =-\langle M(v)|\bar h_v (iD-\delta m)^2 h_v|M(v)\rangle\,,
\ee

\n where the nonrelativistic normalization of states $\langle
M(v)|M(v)\rangle=1$ is implied\footnote{We have changed the sign in
the definition compared to \cite{BAL93}. In the conventional notation
\cite{NEU93} $K=-\lambda_1$.}.
As $\bar{\lambda}$, this matrix element is in fact independent of
$\delta m$.
To derive the sum rule, one considers the correlation function

\be
i^2\int\!dx\!\int\!dy\, e^{i\omega (v\cdot x)-i\omega'(v\cdot y)}
\langle 0| j^\dagger_5(x) \bar h_v(0) (iD-\delta m)^2 h_v (0)
j_5(y)|0\rangle = T_K(\omega,\omega')\,.
\ee

\n Assuming, as before, that the effect of ignoring the
UV renormalon divergence in large orders is equivalent to an
ambiguity in the external momenta (frequencies) and repeating a set of
standard steps we arrive at the sum rule

\bea\label{SRK}
\widehat f_B^2 K_{\delta m}e^{-(\bar\Lambda-\delta m)/\tau} &=&
\frac{3}{\pi^2}\int_0^{\omega_0}\!d\omega\,\omega^4
e^{-\omega/\tau}[1+\mbox{perturbative series}]\nonumber\\
&&\hspace*{-2cm}
+\frac{1}{4}\,\tau\left(1-e^{-\omega_0/\tau}\right)
\langle\frac{\alpha}{\pi} GG\rangle
-\frac{3}{8}\langle g\bar q \sigma Gq\rangle(\mu=2\tau)+\ldots \,.
\eea

\n Again, for simplicity we have discarded the radiative corrections
calculated in \cite{BAL93}.

In eq.(\ref{SRK}) we recognize the familiar source of ambiguity related to
an uncertainty in the position $\bar\Lambda-\delta m$
of the ground state. However, an additional uncertainty is present
already in the definition of the matrix element in eq.(\ref{kinetic})
due to quadratic and linear UV divergences, cf. ref. \cite{MAI92}.
With respect to quadratic divergences it is interesting to note
that the corresponding IR renormalon in the pole mass at $u=1$
is absent, see eq.(\ref{polemass}). In any case,
the sum rule analysis in
\cite{BAL93} has yielded a relatively large value for $K$, of order
0.6 GeV$^2$, which may indicate that the kinetic energy has a large
``genuinely nonperturbative'' contribution, not related to renormalons,
and in this respect is similar to the gluon condensate.

To summarize, the QCD sum rule approach faces precisely
the same difficulties in defining the observables of HQET, which have been
recognized in studies of HQET on the lattice. However, there is also a
difference. In lattice calculations one does not distinguish
betwen perturbative and nonperturbative contributions to the
correlation functions. Thus the renormalon problem is difficult to
overcome, see \cite{MAI92}. In QCD sum rules one isolates the ``genuinely
nonperturbative'' contributions in a few parameters, the vacuum
condensates, which are determined from phenomenology. In spite
of the fact that such an approach can not be fully consistent
theoretically -- the condensates can never be determined to
arbitrary accuracy without running into the renormalon problem
or without the introduction of a ``hard'' factorization scale -- it may
nevertheless be quite successful phenomenologically, as it has been
in the application to light quarks. A novel feature of the QCD
sum rules in HQET is that the ambiguity in the separation of the
perturbative and nonperturbative contributions affects not only the
values of condensates on the ``theoretical" side of the sum rule,
but also the quantities that enter the ``phenomenological" side .
In the HQET, the l.h.s. of sumrules
like eq.(\ref{SRK}) is only fully defined after one has dealt with
the UV renormalons in the perturbative expansions on the r.h.s., though
in practice one might hope to be as lucky as in QCD, where the renormalons
can be ignored, since the ``true'' nonperturbative contributions
to theoretically ambiguous quantities turn out to be large. Thus,
for the practitioner, the appearance of an ambiguous residual mass of
order $(100-200)\,\mbox{MeV}$, see eq.(\ref{poleambiguity}), can serve
as an error bar on the determination of quantities like $\bar{\Lambda}$.

\mysection{Conclusions}{Conclusions}

The investigation of the asymptotic behaviour of perturbative
expansions in HQET reveals that in addition to the IR renormalon
divergence, which can be related to condensates in the SDE, the
correlation functions possess non-summable UV renormalons. These UV
renormalons are not related to a Landau ghost as familiar from
QED, but rather indicate a fine-tuning problem of HQET. The natural mass
of the effective heavy quark is $m_Q$ (and not zero), a fact that
is obscured by the use of dimensional regularization, which does
not introduce a dimensionful parameter, as long as poles at $d=4$ only
are subtracted. The UV renormalons reflect a linear divergence
of the self-energy of the heavy quark, which is seen already in perturbation
theory, when a dimensionful cutoff is employed. The absence of a dimensionful
quantity in the leading effective Lagrangian of HQET is fake and if
one attempts to go beyond perturbation theory, the residual mass
arises necessarily as a second parameter in the Lagrangian. This
is more evident, when one does not consider HQET as a quantum field theory
in its own right, but embedded in QCD, whose heavy mass limit it
is supposed to extract. To avoid the UV renormalon problem, one
can not use the standard dimensional renormalization, which yields
only an incomplete factorization of effects on different distance scales
on the level of logarithms. Technically, complete factorization
can be achieved by a ``hard" cutoff, which is very awkward for
practical calculations. As an alternative, we have indicated a
factorization procedure for renormalons, which on the level of
the HQET Lagrangian corresponds to a mass term proportional to the
scale $\mu$.

The fine-tuning problem of HQET has a very transparent interpretation,
when it is viewed from the perspective of full QCD. HQET (to leading order
in $m_Q$) is a theory for light quarks in the field of a static
colour source. In perturbation theory, this notion does not
present a difficulty. One may imagine the light quark removed to an
infinite distance from the source and include the energy of the
field of the heavy quark into a renormalization of its mass. In this
way, the pole mass emerges naturally as the parameter to be used
in the heavy quark expansion. Beyond perturbation theory, this
operational definition looses its meaning due to confinement. The
meson is an indivisible entity (for QCD) and an unambiguous separation
of an energy of the field and a binding energy of the light quark
in this field can not be performed. Remarkably, perturbation theory
knows about this problem and reveals it as an IR renormalon in the
pole mass. The position of this renormalon in the Borel plane
fixes the inherent ambiguity in the concept of a pole mass to be
of order $\Lambda_{QCD}$. From their physical origin, it is clear
that these IR renormalons are very different from the ones encountered
in the SDE and, in particular, they do not correspond to any
condensate.

Nonetheless, the implications of these IR renormalons for the structure
of the heavy mass expansion are very close conceptually to their
namesakes in the SDE. First, the UV renormalons in the correlation
functions of HQET reflect in fact one and the same phenomenon as the
IR renormalon in the pole mass. If we assume that the Green functions
of QCD can be reconstructed from an extended (and presumably very
intricate) Borel summation procedure -- a conjecture, of course! --
then it is very natural to remedy the ambiguities of correlation
functions in HQET from the UV singularities by the inclusion of an
ambiguous residual mass into the Lagrangian. This leads immediately
to the conclusion that parameters like $\bar{\Lambda}$ (or $\bar{\lambda}$,
to be precise) or the kinetic energy $K$ are not physical quantities,
but in fact ambiguous. This is indeed a necessity, because these
parameters arise in power-suppressed (in $1/m_Q$) terms of the heavy
mass expansion, whose leading term has a divergent perturbative
expansion. In this respect the heavy mass expansion is in complete
analogy with the SDE, where the role of $\bar{\lambda}$ etc. is played
by the condensates, whose theoretically ambiguous status has been realized
a long time ago. In this light the appearance of an IR renormalon
in the pole mass at $t=-1/(2\beta_0)$, which is closer to the origin
of the Borel plane and implies a stronger divergence of perturbative series
than in the SDE, is very natural, since $1/m_Q$-corrections are present in
the heavy quark expansion and are parametrized by $\bar{\lambda}$.

Since the interpretation of the various quantities that appear in
asymptotic expansions with exponentially small (in the coupling) components
such as the SDE or the heavy quark expansion has caused confusions in the
past (see the discussion of this point in refs. \cite{DAV84,NOV84}),
which are merely a problem of language, it
might be useful to recall that there are two attitudes concerning
the renormalon problem, which already have been alluded to in Sect. 2:
First, one can interpret these expansions
as asymptotic expansions in the mathematical sense\footnote{Assuming,
of course, that they {\it are} asymptotic to something.}. Then one faces
the problem of divergent series, their summation and the Stokes
discontinuities in the exponentially small components, which leads
to the notion of formally ambiguous parameters. Second, one might
follow Wilson's operator product expansion literally and introduce a
hard factorization scale $\mu$. In this way, the divergence of perturbative
expansions is eliminated at the price of parameters that depend
explicitly on the scale $\mu$. Both approaches are of course
equivalent in their physical content: The quantities in the power-suppressed
terms are not physical in the sense that they can not be determined to
arbitrary accuracy without further specification. In the first
approach this is a prescription to sum the divergent series in the
leading terms (a principal value prescription, for instance),
in the second, quantities like $\bar{\lambda}$ in the
heavy quark expansion and the gluon condensate in the SDE depend
power-like on the factorization scale. Both approaches are also impractical:
Neither can we sum perturbative expansions
in view of the few low-order terms that are generally available,
nor can we calculate
Feynman diagrams with an explicit cutoff. Thus, although the second
approach looks much more natural to phenomenology, where one is
prepared to fit the unknown quantities anyway, one has to rely in
both cases on the hope that ``true'' nonperturbative contributions
turn out to be large. If nature likes it different, the study of
power-corrections is academic anyway and one should devote oneself
to the calculation of the next unknown order of perturbation theory.

We have chosen the first approach in the present paper because of the
{\it universality} of the UV renormalons in HQET. They arise only
through the linear divergence of the heavy quark self-energy. After
we include the formally ambiguous mass, we can easily trace the
effects of the UV renormalons through the appearance of the residual
mass term in the matrix elements and operators of HQET.

The universality of the phenomenon is also important to
recognize its phenomenological consequences. The inclusive $B$-decay
widths are a prime example of practical interest. The ambiguity of
order $\Lambda_{QCD}$ in the pole mass implies that when parametrized
in terms of the pole mass, the theoretical prediction for the
absolute widths can not be better than terms of order $\Lambda_{QCD}/m_b$.
However, the IR renormalon in the leading term is universal for all
$B$-hadrons and cancels in the difference of the widths, which indeed
scale with the heavy quark mass as $\Lambda_{QCD}^2/m_b^2$.

As a second example, the status of the pole mass itself warrants
discussion. For phenomenology, the most important question is how
large the intrinsic uncertainty of the pole mass could be numerically.
Our estimate from the divergence of the perturbative expansion suggests
values in the range $\delta m_{pole} \sim 170-280$ MeV, but this can
only be an order-of-magnitude guess. There are various indications from
phenomenology that the actual ambiguity is indeed of this order or
rather smaller. All
existing phenomenological analyses of the $b$-quark pole mass fall
in the range 4.55-4.85 GeV, a fraction of
which can well be ascribed to an inherent
ambiguity of the concept ``pole mass". In this context, it is interesting
to note that the existing calculations of the quantity
$\bar\Lambda$ in HQET give $\bar\Lambda = 400-600$
MeV \cite{BAG92,NEU92} with an uncertainty of the same order as
for the pole mass. Last but not
least, it has been pointed out \cite{BJO92}, that a meson with a
light and an infinitely heavy quark might provide a definition of
the constituent quark -- one of the most mysterious objects in
QCD. Indeed, the correlation function of two heavy-light currents,
eq.(\ref{CF2}),
may be rewritten as the vacuum expectation value of the nonlocal
operator \cite{RAD91}

\be
\langle 0|\bar q(x) \mbox{Pexp}[ig\int_0^1 \dd u \,x_\mu A_\mu(ux)]
q(0)|0\rangle
\stackrel{x^2\rightarrow -\infty}{\sim} \exp[-\bar\Lambda \sqrt{-x^2}]
\,,\ee

\n which gives a natural definition of the propagator of a
constituent quark, so that $\bar\Lambda$ may be interpreted as the
constituent quark mass. The celebrated successes of nonrelativistic
quark models (for light quarks) have not found any rational
explanation so far, but indicate rather strongly that the mass of the
constituent quark is phenomenologically stable and of order
350 MeV. This falls into the range of values quoted for $\bar\Lambda$
to $100 - 200\,\mbox{MeV}$ accuracy. Combining these estimates
from different branches of phenomenology, we should conclude that there
is a lot of indirect evidence, that the difference between the hadron mass
and the quark pole mass in the heavy quark limit has a large ``genuinely"
nonperturbative contribution and
the uncertainty of the concept of the pole mass is likely
to stay within $100 -200\,\mbox{MeV}$.\\[0.3cm]

{\bf Acknowledgements.}
It is a pleasure to thank V.I.Zakharov for many interesting discussions
related to the subject of this paper. M.B. wishes to thank M.Einhorn for
an instructive conversation. V.B. gratefully acknowledges discussions
with N.G.Uraltsev, which initiated this study, and our special thanks
are to him for sending us a preliminary version of ref. \cite{BIG94b}.
We acknowledge an overlap with some of the results and conclusions of this
paper.


\newpage
\begin{appendix}

\mysection{Renormalization of the Borel transform}
{Renormalization of the Borel transform}

In this appendix we prove that renormalization of the divergence
associated with the integration over the gluon momentum amounts
to subtracting the pole term of the Borel transform at $u=0$ plus
some finite terms which depend on the renormalization scheme. As
a by-product, we find the anomalous dimension of the heavy quark
field to leading order in $1/N_f$. The derivation is given for the
simplest case of the heavy quark self-energy, but proceeds almost
identically for a massive quark or the vertex function.

We start {\it ab initio} and compute the regularized coefficient
of the heavy quark self-energy in order $a^{n+1}$. We use
dimensional regularization in $d=4+2\epsilon$ dimensions. A
straightforward calculation of the diagram of Fig. 6a yields

\be\label{regcoeff}
p_n^{reg}(d) = \pref\,v k\,\beta_0^n\,
\frac{1}{(n+1) \eps^{n+1}}\, G(d,(n+1)\eps)\,,
\ee

\n where $\beta_0=1/(6\pi)$, $C_F=4/3$ and $G(d,s)$ is given by

\be
G(d,s) = \left(\frac{1}{4\pi}\right)^s \left(-\frac{2 v k}{\mu}
\right)^{2 s}
\left[-6\eps\frac{\Gamma(-\eps) \Gamma(2+\eps)^2}{\Gamma(4+2\eps)}
\right]^{s/\eps-1} \!\!\!(-2 s) \,(3+2\eps) \,\frac{\Gamma(-1-2 s)
\Gamma(1+s)}
{\Gamma(2+\eps-s)}\,.
\ee

\n For later use we collect the definitions

\be G(d,s) = \sum_{j=0}^\infty G_j(d) s^j\,,\qquad G_0(d) =
\sum_{j=0}^\infty g_j\eps^j
\ee

\n with
\bea\label{functions}
G(d,0)\!&=&\!G_0(d) = -\frac{1}{6} \,(3+2\eps)\, \frac{\Gamma(4+2\eps)}
{\Gamma(1-\eps)\Gamma(2+\eps)^3} = -3 - 4\eps + O(\eps^2)\,,\nonumber\\
G(4,s) \!&=&\! \left(-\frac{2 vk}{\mu}\right)^{2 s} e^{s \,(\gamma_E-5/3-
\ln 4\pi)} \,(-6 s)\, \frac{\Gamma(-1-2 s)\Gamma(1+s)}{\Gamma(2-s)}\,.
\eea

\n $\gamma_E$ denotes the Euler-Mascheroni constant.
 The limits $d\rightarrow 4$ and $s\rightarrow 0$
do not commute in the general case (although they do for the
heavy quark self-energy considered here).

The diagram in Fig. 6a consists of two basic renormalization parts:
The fermion bubble and the diagram itself. The counterterm for a
fermion bubble is given by $-\beta_0 (1/\eps + finite)$. We use first
the minimal subtraction scheme and comment on other renormalization
schemes later. Thus we do not subtract finite terms. Take now the
diagram in Fig. 6b, where $k$ bubbles have been replaced by their
counterterms.
Since the only dependence on $n$ on the r.h.s. of eq.(\ref{regcoeff})
originates from the number of fermion loops, substitute $n\rightarrow
n-k$ and multiply by $(-\beta_0/\eps)^k$ for each of the $k$
counterterms. Finally
account for a combinatorical factor $n!/(k! (n-k)!)$ according to
the number of ways, $k$ bubbles can be picked from the $n$ bubbles
available. As a result of these manipulations a partially
renormalized coefficient, incorporating the renormalization of the
coupling $a$, is obtained:

\bea
p_n^{part.ren}(d) \!\!&=&\!\!\pref\,v k\,\beta_0^n\,
\sum_{k=0}^n\frac{1}{\eps^{n+1}}\frac{(-1)^k}{n+1-k} {n\choose k}
G(d,(n+1-k)\eps)\nonumber\\
\!\!&=&\!\!\pref\,v k\,\beta_0^n\,
\sum_{j=0}^{n+1}\frac{G_j(d)}{\eps^{n+1-j}}\sum_{k=0}^n (-1)^k
{n\choose k} (n+1-k)^{j-1}+O(\eps)
\eea

\n The sum over $j$ is truncated at $n+1$, because we are not
interested in terms that vanish, when $\eps$ is taken to zero. The
sum over $k$ can be taken. It is non-zero only for $j=0$ and
$j=n+1$, a simplification that was first observed in \cite{PAL84}.
Thus

\be\label{partren}
p_n^{part.ren}(d) = \pref\,v k\,\beta_0^n
\left[\frac{(-1)^n}{n+1}\frac{1}{\eps^{n+1}} G_0(d) + n!\,G_{n+1}(4)
+ O(\eps)\right]\,.
\ee

\n In the next step we account for the subtraction for the whole
diagram and then take $\eps=0$. It is gratifying that only the finite
terms of $p_n^{part.ren}(d)$ depend on $v k/\mu$, as it must be. The
renormalized coefficient in the $MS$ scheme is then given by

\be
p_n = \pref\,v k\,\beta_0^n \left[
n!\,G_{n+1}(4) + \frac{(-1)^n}{n+1} g_{n+1}\right]\,.
\ee

\n It is easy to see that the $g_{n}$ do not diverge factorially,
because $G_0(d)$ is analytic at $d=4$. Finally, we go over to
the Borel transform of the renormalized self-energy and find

\bea\label{renormbt}
\tilde{\Sigma}_{eff}(vk,u) &=& \sum_{n=0}^\infty p_n\frac{t^n}{n!}
= \pref\,v k\left[\frac{G(4,-u)-G_0(4)}{(-u)} +
R(u)\right]\\
&&\hspace*{-2.7cm}=\,\,\,
\pref\,v k\left[\left(-\frac{2 v k}{\mu}
\right)^{-2 u} e^{-u \,(\gamma_E-5/3-\ln 4\pi)} \,(-6) \,\frac{
\Gamma(-1+2 u)\Gamma(1-u)}{\Gamma(2+u)} - \frac{3}{u} + R(u)
\right]\,,
\nonumber\eea

\n where

\be\label{finiterens}
R(u) = \sum_{n=0}^\infty \frac{1}{(n+1)!} g_{n+1} u^n =
\frac{3}{u} + \frac{\tilde{G}_0(u)}{u}
\ee

\n is an entire function. $\tilde{G}_0$ denotes the Borel
transform of $G_0$ given in eq.(\ref{functions}).
As promised, the renormalized self-energy
differs from the partially renormalized one, obtained from
insertion of the Borel-transformed gluon propagator,
eq.(\ref{gluonprop}), only by subtraction of the pole at $u=0$ and
scheme-dependent finite terms.

Let us dwell more on the issue of scheme-dependence.
Eq.(\ref{renormbt}) is easily generalized to arbitrary minimal
subtraction-like schemes such as $\overline{MS}$. These schemes
have identical renormalization group functions and differ only by
a global scale change. To the order of the $1/N_f$-expansion
considered here, the difference resides in the finite term $C$ of
the fermion bubble. Observing that in the $MS$ scheme
$C=\gamma_E-5/3-\ln 4\pi$, we obtain eq.(\ref{renormbt}) in an
arbitrary $MS$-like scheme by the replacement

\be
e^{-u \,(\gamma_E-5/3-\ln 4\pi)}\longrightarrow e^{-u C}\,.
\ee

\n The function $R(u)$ is the same for all $MS$-like schemes.
In principle, the finite renormalizations and therefore $R(u)$
can be chosen arbitrarily. In momentum subtraction schemes the
finite terms are factorially divergent and $R(u)$ is no longer
an entire function. This introduces a divergence
into the perturbative expansion of the renormalization group functions,
which hides part of
the divergence of the perturbative series in the definition of
the renormalized parameters of the theory. For this reason the
choice of such schemes is disfavoured, when one considers large
orders in perturbation theory \cite{BEN93a}. We will always assume
that a scheme is chosen, where the counterterms are analytic
in $\alpha$ near $\alpha=0$, in which case the function $R(u)$ is
entire in the Borel plane and the renormalized parameters are
unambiguously defined in terms of the bare parameters. In the
general situation of the self-energy of a massive quark $R$ can
also depend on the mass. We restrict ourselves to mass-independent
schemes.

We conclude this appendix with the anomalous dimension of the
heavy quark field in $MS$-like schemes. The bare field is related
to the renormalized
one through $h_v^0=Z^{1/2} h_v$. From eq.(\ref{partren}) we get

\be
Z^{-1} = 1 + \frac{a(\mu)}{4\pi N_f} C_F\,\sum_{n=0}^\infty (\beta_0
a(\mu))^n\,\frac{(-1)^n}{(n+1)\eps^{n+1}} G_0(d) + finite +
O\left(\frac{1}{N_f^2}\right)\,.
\ee

\n The anomalous dimension of the field is defined by

\be
\gamma(a) \equiv \mu^2\frac{\partial Z}{\partial\mu^2} =
\lim_{\eps\rightarrow 0}\beta(\eps,a) \frac{\partial Z}
{\partial a}\,.
\ee

\n Recalling $\beta(\eps,a)=\eps a+\beta_0 a^2+
O(1/N_f)$, a short calculation yields

\be
\gamma(a) = \frac{a}{4\pi N_f} C_F\,G_0(4-\beta_0 a) + O\left(
\frac{1}{N_f^2}\right)\,.
\ee

\n We remind the reader that the anomalous dimension is
gauge-dependent and the result is given in the Landau-gauge.
However, one may easily check that to order $1/N_f$ only the
$O(a)$-coefficient is gauge-dependent.
As expected in $MS$-like schemes, the anomalous dimension has
a finite radius of convergence in $a$. Comparison with
eq.(\ref{finiterens}) shows that the finite renormalizations
in a given scheme are essentially the Borel transform of the
anomalous dimension.

\mysection{The scalar two-loop intergal}
{The scalar two-loop intergal}

The aim of this appendix is to find an expression for the
scalar integral

\be S(d,\gamma) = \frac{(4\pi)^d}{4\,(2\omega)^{2 d-6-2\gamma}}
\int\frac{\dd^d k}{(2\pi)^d}\frac{\dd^d p}{(2\pi)^d}\,
\left(-\frac{1}{k^2}\right) \left(-\frac{1}{p^2}\right)
\left(-\frac{1}{(k-p)^2}\right)^\gamma \left(\frac{1}{v k+\omega}
\right) \left(\frac{1}{v p+\omega}\right)
\ee

\n with arbitrary complex $\gamma$, which is suited to extract
its analyticity properties. The prefactor is chosen for
later convenience. The first step is to rewrite this integral
in coordinate space and to apply the Gegenbauer polynomial
technique \cite{CHE80}. Following the by now standard procedure,
we arrive at

\be
S(d,\gamma) = 2\,\frac{\Gamma(d/2-1)\Gamma(d/2-\gamma-1)
\Gamma(6+2\gamma-2 d)}{\Gamma(d-2)}\,P(d,\gamma)\,,\ee
\bea P(d,\gamma) &=& \frac{1}{\Gamma(\gamma)}\sum_{n=0}^\infty
\frac{(-1)^n}{n!}\,\Gamma(d-2+n)\nonumber\\
&&\times\,\intl_0^{1/2}\dd u\left(\frac{1}{n+\gamma} u^{2+2\gamma
-d+n}\bar{u}^{2-d-n} + \frac{1}{\gamma-d-n+2} u^n\bar{u}^{2\gamma
-2 d+4-n}\right)
\eea

\n with $\bar{u}\equiv 1-u$. The summation can be taken at the
price of another integration and yields

\be
P(d,\gamma) = \frac{\Gamma(d-2)}{\Gamma(\gamma)}\intl_0^{1/2}
\dd u\intl_0^1\dd t\,\frac{t^{d-3-\gamma}}{(1-u\bar{t})^{d-2}}
\left\{(t u)^{2-d+2\gamma}-\bar{u}^{2-d+2\gamma}\right\}\,.
\ee

\n Next, the denominator is expanded, which allows to perform the
$t$-integration:

\bea
P(d,\gamma) &=& \sum_{n=0}^\infty\frac{1}{3-d+2\gamma+n}
\left(\frac{1}{2}\right)^{3-d+2\gamma+n}\frac{\Gamma(d-2+n)}
{\Gamma(n+\gamma+1)}\nonumber\\
&&-\,\frac{\Gamma(d-2)}{(d-2-\gamma)\Gamma(\gamma)}\intl_0^{1/2}
\dd u\,\bar{u}^{2-d+2\gamma}{}_2 F_1(d-2,1,d-1-\gamma;u)\,,
\eea

\n where ${}_2 F_1$ is the hypergeometric function. Now we
do some juggling with the hypergeometric function in order
to replace the hypergeometric function in this equation by
hypergeometric functions with the argument $\bar{u}$. When
this is done, part of the $u$-integrals can be done. Further,
we note that ${}_2 F_1(1-\gamma,d-2-\gamma,1-\gamma;\bar{u})
= u^{2-d+\gamma}$ and obtain

\be
P(d,\gamma) = \sum_{n=0}^\infty\frac{1}{3-d+2\gamma+n}
\frac{\Gamma(d-2+n)}{\Gamma(n+\gamma+1)} - \Gamma(d-2-\gamma)
\intl_0^{1/2}\dd u\,(u\bar{u})^{2-d+\gamma}\,.
\ee

\n The integral is elementary and the sum can
be expressed in terms of the generalized hypergeometric
function ${}_3F_2$. Thus

\be\label{aa}
P(d,\gamma) = -\frac{1}{2}\frac{\Gamma(1-s)\Gamma(s)^2}{\Gamma(2 s)}
+ \frac{\Gamma(\gamma+1-s)}{(s+\gamma) \Gamma(\gamma+1)}\,
{}_3 F_2(\gamma+1-s,1,s+\gamma;\gamma+1,\gamma+1+s;1)\,,
\ee

\n where $s\equiv 3-d+\gamma$ and the series definition of the
${}_3 F_2$-function is

\be
{}_3 F_2(a,b,c;\alpha,\beta;z) = \sum_{n=0}^\infty
\frac{\Gamma(a+n)}{\Gamma(a)} \frac{\Gamma(b+n)}{\Gamma(b)}
\frac{\Gamma(c+n)}{\Gamma(c)} \frac{\Gamma(\alpha)}{\Gamma(\alpha+n)}
\frac{\Gamma(\beta)}{\Gamma(\beta+n)}\,\frac{z^n}{n!}\,.
\ee

\n The series, when applied to eq.(\ref{aa}), converges for
$\gamma>d-3$ only, which is not yet
sufficient for our purpose. We therefore use the identity \cite{PRU86}

\be
{}_3F_2(\gamma+1-s,1,s+\gamma;\gamma+1,\gamma+1+s;1) =
\frac{\Gamma(\gamma+1)\Gamma(\gamma+1+s)}
{s\,\Gamma(\gamma+1-s)\Gamma(2 s+\gamma)}\,
{}_3 F_2 (s,2 s,s;s+1,2 s+\gamma;1)\,,
\ee

\n which provides a representation that converges for any $\gamma$,
as long as $d>2$. The final answer for $S(d,\gamma)$ is

\bea\label{scalarintegral}
S(d,\gamma) &=& \frac{\Gamma(d/2-1)\Gamma(d/2-\gamma-1)}{\Gamma(d-2)}
\Bigg\{-\Gamma(1-s)\Gamma(s)^2 \nonumber\\
&&\hspace*{0cm} + 2\,\frac{\Gamma(s+\gamma)\Gamma(2 s)}
{s\,\Gamma(2 s+\gamma)}\,{}_3 F_2(s,s,2 s;s+1,2 s+\gamma;1)
\Bigg\}\,.
\eea

\n In four dimensions $S(4,\gamma)$ is meromorphic with poles
at all integers, negative half-integers and $\gamma=1/2$.
Useful expansions about some of the singularities in four
dimensions are:

\bea
&&\hspace*{-2.5cm}
S(4,-1/2+\delta) = -\frac{1}{3\delta^2}+O\left(\frac{1}{\delta}
\right)\qquad
S(4,\delta) = \frac{1}{2\delta^2}+\frac{1}{2\delta} +  O(1)\nonumber\\
&&\hspace*{-2.5cm}
S(4,1/2+\delta) = -\frac{1}{\delta^2}+O\left(\frac{1}
{\delta}\right)\qquad\hspace*{0.5cm}
S(4,1+\delta) = \frac{\pi^2}{3\delta} +  O(1)\nonumber\\
&&\hspace*{-2.5cm}
S(4,2+\delta) = \frac{1}{\delta^2}+O(1)\qquad \hspace*{1.6cm}
S(4,3+\delta) = \frac{1}{2\delta^2} - \frac{2}{3\delta}
+ O(1)\nonumber\\
&&\hspace*{-2.5cm}
S(4,4+\delta) = \frac{1}{3\delta^2} + \frac{3}{10\delta} + O(1)
\eea

\end{appendix}



\newpage
\normalsize
\begin{center}
{\large\bf Figure Captions}
\end{center}
\vspace*{1cm}

\n {\bf Fig.1} Two-loop diagram for the correlation function of
vector currents.\\[0.1cm]

\n {\bf Fig.2} Definition of the Borel-transformed gluon propagator.
Renormalization of the fermion loops is implied.\\[0.1cm]

\n {\bf Fig.3} Diagram for the Borel transform of the self-energy of
a heavy quark in HQET.\\[0.1cm]

\n {\bf Fig.4} Diagrams for the Borel transform of the correlation
function of heavy-light currents in HQET. The shaded circle stands
for an insertion of the current with momentum $q$, the double line
denotes the effective heavy quark propagator.\\[0.1cm]

\n {\bf Fig.5} Singularities in the Borel plane to order $1/N_f$ for
(a) the correlation function of vector currents in QCD and (b) the
correlation function of heavy-light currents in HQET.\\[0.1cm]

\n {\bf Fig.6} (a) Diagram with $n$ fermion loops contributing to
the self-energy to order $a^{n+1}$. (b) Counter-diagram with $k$
bubbles replaced by counterterms.


\end{document}